\documentclass[11pt, reqno]{article}
\usepackage{jheppub}
\usepackage{graphicx}
\usepackage{amsmath}
\usepackage{amssymb}
\usepackage{amsthm}
\usepackage{tikz-network}
\usepackage{mathrsfs}
\usepackage{hyperref}
\usepackage{xcolor}
\usepackage{subfigure}
\usepackage{booktabs} % For professional-looking horizontal lines
\newcommand{\nn}{\nonumber \\}

\def\newline{{\hspace{15pt}}}

\def\abs#1{\left| #1\right|}
\def\braket#1{\left\langle #1 \right\rangle}

\def\wt#1{ \mathsf{w} #1}

\def\dd{\text{d}}

\def\what{\widehat}
\def\co{\,,}
\def\ed{\,.}

\title{\boldmath Higher-Order Corrections to Scrambling Dynamics in Brownian Spin SYK Models  }

\author[a,b,c]{Tingfei Li,}
\author[d]{Miao Wang,}
\author[d]{Jianghui Yu}

\affiliation[a]{College of Physics Science and Technology, Hebei University, Baoding, 071002, China}
\affiliation[b]{Hebei Key Laboratory of High-precision Computation and Application of Quantum Field Theory, Baoding, 071002, China}
\affiliation[c]{Hebei Research Center of the Basic Discipline for Computational Physics, Baoding, 071002, China}
\affiliation[d]{Kavli Institute for Theoretical Sciences (KITS), University of Chinese Academy of Sciences, Beijing 100190, China}

\emailAdd{tfli@zju.edu.cn}
\emailAdd{wangmiao21@mails.ucas.ac.cn}
\emailAdd{yujianghui21@mails.ucas.ac.cn}

\abstract{
	We investigate operator growth in a Brownian spin Sachdev--Ye--Kitaev (SYK) model with random all-to-all interactions, focusing on the full operator-size distribution. For Hamiltonians containing $q$-body interactions, we derive a closed master equation for the Pauli-string expansion coefficients and recast their dynamics into a generating-function formulation suitable for the large-\(N\) limit. This approach allows us to diagonalize the leading-order evolution operator explicitly and obtain exact solutions for arbitrary initial operator distributions, including the effects of decoherence. Going beyond leading order, we develop a systematic \(1/N\) expansion that captures higher-order corrections to the operator-size dynamics and the late-time behavior. 
	Our results demonstrate that higher-order effects play a crucial role in operator scrambling and that the full operator-size distribution provides a more refined probe of quantum chaos in Brownian and open quantum systems.
}


\begin{document}
	\maketitle
	\flushbottom
	\section{Introduction}
	\paragraph{Background} 
	Scrambling refers to the dynamical process in quantum many-body systems whereby initially localized information becomes highly nonlocal, encoded in increasingly complex correlations. Consequently, this information becomes effectively inaccessible to local measurements despite being preserved under unitary time evolution. A central quantitative probe of scrambling~\cite{Lewis-Swan:2019xbi, Xu_2024} is provided by \emph{out-of-time-order correlators} (OTOCs)~\cite{Larkin1969QuasiclassicalMI}, which diagnose the growth of noncommutativity between initially simple operators. Beyond their foundational role, OTOCs have been linked to entanglement entropy~\cite{FAN2017707} and have been experimentally measured on various platforms~\cite{Li_2017, PhysRevLett.120.070501, PhysRevA.100.013623, Gu_2022, Li_2024}. Microscopically, this process is described as \emph{operator growth}: in the Heisenberg picture, a simple operator spreads into a superposition of increasingly nonlocal operators~\cite{Nahum_2018, vonKeyserlingk2018, Parker_2019}, reflecting the delocalization of quantum information across the system. While operator growth is dual to entanglement generation in the Schrödinger picture~\cite{Kim:2013etb,Ho_2017,Nahum_2017,swann2023}, the operator-size distribution offers a more fine-grained probe of scrambling dynamics than the mean size alone. It captures the full spectrum of operator growth during unitary time evolution~\cite{Roberts_2018, Qi_2019, Lucas_2020, Zhang_2023} as well as in open quantum systems~\cite{Zhangoperator_2023, Schuster2023}, and has also been addressed in experimental studies~\cite{qi2019measuringoperatorsizegrowth}.

	In realistic experimental settings, however, probing scrambling is complicated by the presence of noise and experimental imperfections. While unitary evolution delocalizes information via operator growth, decoherence and imperfect time reversal can obscure or even mimic the decay of OTOCs expected from intrinsic unitary dynamics. Motivated by this challenge, \cite{Nadie2025} introduced a protocol based on \textit{dressed} OTOCs and an echo observable, which explicitly incorporates unequal \textit{forward} and \textit{backward} evolutions as well as depolarizing noise.\footnote{The influence of noise on OTOCs and quantum chaos diagnostics has also been examined in \cite{Li_2026} and \cite{Li:2025gil}, respectively.} By studying the ratio between the dressed OTOC and the echo signal (termed the renormalized OTOC, or ROTOC, and defined in \cite{PhysRevA.97.062113}), that work provided a framework for disentangling genuine scrambling dynamics from experimental imperfections.

	More concisely, \cite{Nadie2025} analyzed this protocol in the context of a Brownian spin Sachdev–Ye–Kitaev (SYK) model—an all-to-all interacting Brownian circuit defined on $N$ qubits with a time-dependent (forward) Hamiltonian
	\begin{align}
		H(t) = \sum_{i<j, \alpha \beta}^N J_{ij}^{\alpha \beta}(t) \sigma_i^{\alpha} \sigma_j^{\beta} \equiv \sum_A J_A O_A
	\end{align}
	where \( i, j \) label sites, \( \alpha, \beta \in \{1,2,3\} \) label Pauli indices (with \( \sigma^1, \sigma^2, \sigma^3 \) corresponding to \( X, Y, Z \)), and \( J_A(t) \) are independent Gaussian random variables with zero mean and covariance
	\begin{align}
		\mathbb{E}[J_A(t) J_{A'}(t')] = \mu_2^2 \delta_{AA'} \delta(t-t')\ed
	\end{align}
	Here $\mathbb{E}$ denotes the ensemble average. 
	The Brownian spin SYK model is a type of quantum many-body system known as Brownian models~\cite{Lashkari2013,saad2019semiclassicalrampsykgravity,Jian_2021,Stanford_2022}, characterized by its inclusion of time-dependent, stochastic couplings that act as a source of dephasing noise. A similar model is also discussed in~\cite{Erd_s_2014,Baldwin_2020,Berkooz_2018, Sunderhauf:2019djv,Yin:2020pjd,Swingle:2023nvv,Hanada_2024,anschuetz2024boundsgroundstateenergy,Xu_2025,basu2025complexityquadraticquantumchaos} as spin SYK model which served as an extension of the conventional SYK model made by fermions~\cite{Kitaev2015,Polchinski_2016,Maldacena_2016,Jevicki:2016bwu,jevicki2016bilocalholographysykmodel}.

	In the Brownian spin SYK model, ensemble-averaged correlation functions such as OTOCs can be expressed entirely in terms of the operator-size distribution. The simplification afforded by ensemble averaging permits the derivation of a master equation governing its time evolution:
	\begin{align}\label{eq:bw_eq}
		\frac{d}{dt} \boldsymbol{b}(t) = M.\boldsymbol{b}(t)\ed
	\end{align}
	Here $\boldsymbol{b}(t)=(b_1(t),b_2(t),\ldots,b_N(t))^T$ denotes the operator-size distribution vector, where $b_w(t)$ collects the total contribution of Pauli strings with weight $w$. The matrix $M$ is the transition-rate generator in weight space: its element $M_{ww'}$ gives the ensemble-averaged rate at which an operator component of weight $w'$ contributes to weight $w$ under the Brownian evolution, including the effects of imperfect time reversal and decoherence when present. The weight-zero sector corresponds to the identity operator, whose evolution is trivial and does not mix with operators of nonzero weight, and is therefore omitted.  Although this linear equation can in principle be solved by diagonalizing the $N \times N$ matrix $M$, the task grows prohibitively difficult as $N$ increases, and a general analytic solution remains elusive in this formulation. \cite{Nadie2025} therefore focused on the dilute limit, where the typical operator size $w$ satisfies $w \ll N$.\footnote{We emphasize that the term ``dilute limit'' in this work refers to the regime where a systematic large-$N$ expansion is performed, \textit{not} to the strict mathematical limit $N \to \infty$. 
		Therefore, when analyzing late-time behavior, $N$ is held fixed at a large but finite value, and the limit $t \to \infty$ is taken subsequently.} In this limit, the transition matrix $M$ becomes lower triangular at leading order, allowing for an explicit solution for a restricted class of initial operator distributions. This yields a closed-form expression for the dressed OTOC (and echo) at leading order in the $1/N$ expansion.
	
	However, this triangular structure relies crucially on retaining only the leading-order contributions. Once higher-order corrections are included, $M$ generically loses this simplifying property, making the analytical treatment of late-time operator growth for general initial conditions substantially more challenging. Moreover, except for special initial conditions, the leading-order description obscures the physical mechanisms that govern late-time operator growth.

	\paragraph{Main method} In this work, we develop a systematic extension of the Brownian circuit framework that \textbf{incorporates} higher-order corrections within the dilute limit. We generalize the model to include arbitrary $q$-body interactions, including mixtures of different interaction types, and introduce a \textit{generating-function method} that recasts the operator-growth problem into the solution of a partial differential equation. This approach avoids explicit matrix diagonalization and enables a controlled perturbative expansion in $1/N$. As a result, it provides analytical access to operator dynamics for arbitrary initial distributions, including operators initially localized at a fixed weight $w = m$, for which corrections up to order $m-1$ are generally required.
	
	Our approach is formulated in the large-$N$ limit. We first extend the dimension of the original matrix $M$ formally to infinity, obtaining $M_\infty$, whose entries retain their explicit dependence on the system parameter $N$. We then define an \textbf{approximate} generating function $\mathsf{G}(x, t)$ by \footnote{In Appendix of \cite{Nadie2025}, the author employs the generating function technique to derive the leading-order results. Meanwhile, in \cite{Zhang:2022fma}, the authors first consider the large $N$ limit and subsequently take the continuous limit, after which the generating function method is introduced.  }
	\begin{align}\label{eq:main_approx}
		\mathsf{G}(x,t) = X_\infty \cdot e^{M_\infty t} \boldsymbol{b}_\infty(0) \approx \sum_{w=1}^N b_w(t) \, x^w = X_N \cdot e^{M t} \boldsymbol{b}(0)
	\end{align}
	where $X_\infty = [x, x^2, \dots]$ and $\boldsymbol{b}_\infty(0)$ is the natural extension of the initial operator-size distribution obtained by appending zeros for $w > N$. Although the physical weight distribution cannot exceed $N$, approximating the finite-dimensional matrix $M$ by an infinite-dimensional counterpart remains an excellent approximation when the average operator size is much smaller than $N$.
	From Eq.~\eqref{eq:main_approx}, a partial differential equation for $\mathsf{G}(x,t)$ follows directly
	\begin{align}
		\partial_t \mathsf{G}(x,t)= \mathcal{M}_x \mathsf{G}(x,t)\ed
	\end{align} 
	Here, \(  \mathcal{M}_x \) is a differential operator with respect to the auxiliary variable \( x \). In a certain sense, this partial differential equation resembles the Schrödinger equation, and its solution reduces to solving the eigenvalue problem of \(  \mathcal{M}_x \). In the large-$N$ limit, we can expand \(  \mathcal{M}_x \) in powers of \( 1/N \) and then treat the eigenvalue problem perturbatively. The only formal difference from a standard quantum mechanical problem is that \(  \mathcal{M}_x \) is not a Hermitian operator.
	\paragraph{Main results} 
	At leading order, we find the eigenfunctions obey a power-law structure
	\begin{align}
		G_k^{(0)}(x)=\left(G_1^{(0)}(x)\right)^k .
	\end{align}
	For two- and three-body interactions, the higher-order corrections (up to second order) can be obtained by acting with differential operators on $G_k^{(0)}(x)$
	\begin{align}
		G^{(n)}_k(x)=\widehat{O}^{(n)}_x G_k^{(0)}(x)\ed
	\end{align}
	This ultimately yields a closed expression for the time-dependent generating function valid for any initial operator-size distribution:
	\begin{align}
		\mathsf{G}^{(n)}(x,t)=\widehat{D}_x^{(n)}\mathsf{G}_{\text{init}}(x_t)+\widehat{K}_x^{(n)}\mathsf{G}_{\text{init}}(x)\big\vert_{x\to x_t}
	\end{align}
	where $x_t$ is defined implicitly by
	\begin{align}
		G_1^{(0)}(x_t)=e^{\lambda_1^{(0)}t}G_1^{(0)}(x)\ed
	\end{align}
	Beyond its technical utility, the generating-function approach unveils new physical structure in the operator-growth dynamics that is hidden at leading order. Higher-order corrections induce systematic mixing among dynamical modes associated with different operator weights, producing a hierarchical perturbative structure that is crucial for determining the late-time behavior.

	We illustrate these effects through explicit calculations for two-body and three-body interactions up to second order in the $1/N$ expansion. Excellent agreement with numerical simulations is found, especially in the late-time regime, confirming that higher-order corrections are essential for capturing the complete temporal evolution of operator growth within the dilute limit. More broadly, our results provide a systematic and physically transparent framework for understanding scrambling dynamics in Brownian circuit models in the presence of experimental imperfections. We also expect that the generating-function method developed here may be applicable to the calculation of Krylov complexity~\cite{Parker_2019, Xu:2019lhc, rabinovici2025krylovcomplexity, Gamayun_2025} beyond known integrable cases.

	\paragraph{Structure of the paper} The remainder of the paper is structured as follows. Section~\ref{sec:setup} introduces the generalized Brownian spin model and defines the dressed OTOC and echo observables. Section~\ref{sec:eq} derives the master equation for the operator-size distribution and introduces the generating function method for solving the model in the dilute limit (to leading order in $1/N$). Section~\ref{sec:L2} details the perturbative treatment up to second order, using two-body interactions as a concrete example, and Section~\ref{sec:L3} presents the main analytical results for three-body interactions. In both sections, numerical comparisons are provided to validate the analytical framework. We conclude with a discussion of the physical implications and future directions in Section~\ref{sec:discussion}.
	
	\section{Model and Setup}\label{sec:setup}
	In this section, we provide a concise review of the basic definition of the Brownian SYK model, along with the dressed OTOC and R-OTOC as proposed in the literature. Additionally, we discuss the operator size distribution in this model and its relation to quantities such as the OTOCs. While these topics can be found in similar research articles (like \cite{Xu_2025}), they are included here for completeness and coherence. Throughout the discussion, we also present the main results of this paper.
	\subsection{Brownian spin SYK model and operator dynamics}
	In this work, we study operator growth in a generalized Brownian spin model, which extends the Brownian spin SYK–type constructions introduced in~\cite{Nadie2025,Lashkari_2013,Xu_2025}. The system consists of $N$ qubits evolving under a time-dependent, stochastic Hamiltonian with all-to-all interactions. We allow for interactions of arbitrary order, including mixed interaction types, so that the Hamiltonian takes the schematic form
	\begin{align}\label{eq:H}
		H=\sum_{n=2}^L\sum_{A_n}J_{A_n}(t)O_{A_n}\,,
	\end{align}
	where $O_{A_n}$ denotes an $n$-body operator acting on the subset of sites $A_n$. Here $A_n=\{j_1,j_2,\ldots,j_n\}$ and
	\begin{align}
		O_{A_n}=\bigotimes_{k=1}^n \sigma_{j_k},\sigma_{j_k} \in \{X,Y,Z\}\,.
	\end{align}
	The couplings $J_{A_n}(t)$ are independent Gaussian random variables with
	\begin{equation}
		\mathbb{E}(J_{A_n}(t)J_{A_m'}(t')) =\delta_{A_nA_m'}\delta(t-t')\mu_n^2\,.
	\end{equation}
	Given a realization of the Hamiltonian, an operator $O(t)$ evolves in the Heisenberg picture as
	\begin{align}
		O(t) = e^{iHt} O(0) e^{-iHt}\,.
	\end{align}
	Because the dynamics are Brownian, the physical observables of interest must be obtained by averaging over an ensemble of Hamiltonian realizations. In this work, we concentrate on ensemble-averaged quantities that characterize operator growth and information scrambling.
	
	Here, we temporarily disregard decoherence and experimental imperfections, so a sample realization of the model is equivalent to a conventional quantum mechanical system. Under time evolution, a simple operator can grow increasingly complex. This growth can be quantified using measures such as \textit{Krylov complexity} and \textit{operator size}, which provide effective diagnostics for quantum chaos. While Krylov complexity offers a universal framework for general quantum systems, the operator size and its distribution are particularly well-suited for systems of qubits. In this paper, we focus on the dynamics of the operator size distribution. A convenient basis for describing operator dynamics is provided by Pauli strings, since any operator can be expanded in this basis. A \textbf{Pauli string} of length \( N \) is a tensor product of \( N \) operators, where each operator is chosen from the four \( 2 \times 2 \) Hermitian basis matrices: the identity matrix \( I \) and the three Pauli matrices \( X \), \( Y \), and \( Z \):
	\begin{align}
		P = \bigotimes_{i=1}^{N} \widetilde{\sigma}_i, \quad \text{where } \widetilde{\sigma}_i \in \{ I, X, Y, Z \}\ed
	\end{align}
	The \textbf{weight} $\wt(P)$ of a Pauli string is defined as the number of non-identity operators ($X,Y,$ or $Z$) appearing in its tensor product. Then we can expand any operator in the system into Pauli string basis
	\begin{align}
		O(t)=\sum_P c_P(t)P,~c_P(t)=(P|O(t))\co
	\end{align}
	where we have defined the inner product of two operators 
	\begin{align}
		(O_1|O_2) \equiv  {\text{Tr}(O_1^\dagger O_2)\over \text{Tr}{1}}\,.
	\end{align}
	Here, we define the \textit{average weight} of the operator as a measure of complexity, which is given by: 
	\begin{align}
		\wt(O(t))\equiv \sum_P w_P \abs{c_P(t)}^2=\sum_{w=1}^\infty wb^{\text{clean}}_w 
	\end{align}
	where we collect the contribution of Pauli strings with the same weight:
	\begin{align}
		b^{\text{clean}}_w=\sum_{\wt(P)=w} |c_P(t)|^2\,.
	\end{align} 
	Under unitary evolution, this distribution is normalized $\sum_w b^{\text{clean}}_w(t) = 1$, reflecting the preservation of the operator's trace.

	\subsection{Construct ROTOC with operator size distribution $b_w$}
	
	A major motivation for studying Brownian circuit models comes from their relevance to experimental probes of quantum scrambling. In realistic experiments, \textit{imperfections} such as control errors and \textit{decoherence} can obscure ideal scrambling dynamics. To address this, \cite{Nadie2025} introduced a class of observables known as dressed out-of-time-order correlator (dressed OTOC), defined through protocols involving unequal \textit{forward} and \textit{backward} time evolutions.
	
	In particular, \cite{Nadie2025} considered a perturbed backward evolution generated by a Hamiltonian $\widetilde{H}(t)$, whose couplings are correlated with those of $H(t)$ through a parameter $r$ that quantifies experimental imperfections. Within this framework, they defined the echo and dressed OTOC, and introduced the circuit-averaged ``renormalized OTOC" (ROTOC) as a diagnostic that isolates intrinsic scrambling dynamics from experimental noise.

	Let's consider the case where the forward and backward time evolution are governed by distinct Hamiltonians, denoted as $H$ (same as Eq.~\eqref{eq:H}) and $\widetilde{H}$:
	\begin{equation}\label{eq:sysH}
		\begin{aligned}
			&H=\sum_{n=2}^L\sum_{\wt(O_A)=n} J_A(t)O_A\;,\ \ \mathbb{E}(J_A(t)J_B(t'))=\delta_{AB}\delta(t-t')\mu_{\wt(O_A)}^2\,;\\
			&\widetilde{H}=\sum_{n=2}^L\sum_{\wt(O_A)=n} \widetilde{J}_A(t)O_A\;,\ \ \mathbb{E}(\widetilde{J}_A(t)\widetilde{J}_B(t'))=\delta_{AB}\delta(t-t')\mu_{\wt(O_A)}^2\,;\\
			&\mathbb{E}(J_A(t)\widetilde{J}_B(t'))=r\delta_{AB}\delta(t-t')\mu_{\wt(O_A)}^2\,.
		\end{aligned}
	\end{equation}
	The correlation between $H$ and $\widetilde{H}$ is characterized by the parameter $0\le r\le 1$. In general cases the experimental imperfection causes $r<1$, while the case $r=1$ corresponds to identical forward and backward time evolution. We define the time evolution of an operator by the perturbed Hamiltonian for a single realization as 
	\begin{align}\label{eq:tildeO}
		\widetilde{O}(t)= e^{i\widetilde{H}t}O(0)e^{-i\widetilde{H}t}\ed
	\end{align}
	In this paper, since we focus on ensemble-averaged dynamics and incorporate experimental imperfections, we define $b_w$ as 
	\begin{align}
		b_w\equiv \sum_{\wt(P)=w} \mathbb{E}(\widetilde{c}_P(t)c_P(t))
	\end{align}
	where $c_P(t)$ and $\widetilde{c}_P(t)$ denote the Pauli-string coefficients of the forward- and backward-evolved operators, respectively, and $\mathbb{E}[\cdot]$ denotes the ensemble average. The quantities $b_w(t)$ encode the full information content of operator growth in the presence of imperfections.
	
	We now incorporate \textit{decoherence effects} by introducing a depolarizing channel acting at rate $\kappa$.\footnote{See Appendix A of~\cite{Nadie2025} for detailed discussion about the decoherence.} For the density matrix of the system
	\begin{align}
		\rho(t) &= \sum_{P} f_P(t)P\co
	\end{align}
	after an infinitesimal time $\delta t$, the decoherence effect leads to a decay of the off-diagonal elements in $\rho(t)$
	\begin{align}
		\delta_D\rho(t)&=\sum_{P}(1-\kappa\delta t)^{\wt(P)}f_P(t)P-\rho(t)=-\kappa\sum_{P} \wt(P)f_P(t)P\delta t+\mathcal{O}(\delta t^2)\ed
	\end{align}
	So in the Schrödinger picture, the total (forward) time evolution of the density matrix is given by
	\begin{align}
		\delta \rho(t)=\delta_H\rho(t)+ \delta_{D}\rho(t)\ed
	\end{align} 
	Here $\delta_H\rho(t)\equiv e^{-iH_{t}\delta t}\rho(t)e^{iH_{t}\delta t}-\rho(t)$. 
	It is equivalent to define the total time evolution for an operator $O(t)=\sum_P c_P(t)P$:
	\begin{align}
		\delta O(t)= \delta_H^*O(t)- \kappa\sum_{P} \wt(P)c_P(t)P\delta t+\mathcal{O}(\delta t^2)
	\end{align}
	where we define $\delta_H^*O(t)\equiv e^{iH_{t}\delta t}O(t)e^{-iH_{t}\delta t}-O(t)$. Due to the effects of imperfections and decoherence, generally, we find $\sum_w b_w$ is no longer conserved. 
	So we define the normalized distribution $c_w$ and the average of any function $f(w)$ of weight as 
	\begin{align}
		c_w={b_w\over \sum_{k=1}^{\infty} b_k}\;,\;	\langle f \rangle_c \equiv \sum_w f(w)c_w\,.
	\end{align} 
	The dressed echo and OTOC is defined as 
	\begin{align}
		\widetilde{C}_{W}(t)=\braket{\widetilde{W}(t)W(t)},~\widetilde{C}_{W,V}(t)=\braket{\left[\widetilde{W}(t),V\right]^{\dagger}\Big[W(t),V\Big]},
	\end{align}
	where $\langle \bullet \rangle \equiv \operatorname{Tr}(\bullet)/\operatorname{Tr} 1$. Here, $\widetilde{W}(t)$ and $W(t)$ denote the time-evolved operators starting from the same initial operator $W$, but evolving under the Hamiltonians $\widetilde{H}$ and $H$, respectively. As before, we expand both $\widetilde{W}(t)$ and $W(t)$ in the Pauli basis: $\widetilde{W}(t)=\sum_{P}\widetilde{c}_P(t) P$,$W(t)=\sum_{P}c_P(t) P$. Using the properties of Pauli basis and the couplings, the ensemble averaged dressed echo and OTOC can be simplified to
	\begin{align}
		\text{Echo}_W=\mathbb{E}\left[\widetilde{C}_{W}(t)\right]=4\sum_{w}b_{w},~\mathbb{E}\left[\widetilde{C}_{W,V}(t)\right]=\sum_{P|\{P,V\}=0}4\mathbb{E}\left[c_P(t)\widetilde{c}_P(t)\right]\ed
	\end{align}
	One can find the dressed OTOC can be expressed in terms of $b_w(t)$ as follows
	\begin{align}
		\mathbb{E}\left[\widetilde{C}_{W,V}(t)\right]=\sum_w\frac{4C_{w}^{\wt(V)}}{N_w}b_w(t)\ed
		\label{eq:dressedOTOC}
	\end{align}
	Here $C_{w}^{\wt(V)}$ (see Eq.~\eqref{eq:C_const}) denotes the number of weight $w$ Pauli strings anticommuting with the fixed Pauli string $V$ and $N_w=3^{w}\binom{N}{w}$ is the total number of weight $w$ Pauli strings.  For different weights $\wt{(V)}$, we have
	\begin{equation}
		\begin{aligned}
			&\mathbb{E}\left[\widetilde{C}_{W,V}(t)\right]\Big|_{\wt{(V)}=1}=\sum_w\frac{8 w}{3 N}b_w\,,\\
			& \mathbb{E}\left[\widetilde{C}_{W,V}(t)\right]\Big|_{\wt{(V)}=2}=\sum_w\frac{16 w (3 N-2 w-1)}{9 (N-1) N}b_w\,, \\
			&\mathbb{E}\left[\widetilde{C}_{W,V}(t)\right]\Big|_{\wt{(V)}=3}=\sum_w\frac{8 w \left(9 N^2-3 N (4 w+5)+4 w^2+12
				w+2\right)}{9 (N-2) (N-1) N} b_w\,.
		\end{aligned}
	\end{equation}
	To study scrambling, we are more interested in the ratio of the OTOC to echo. The renormalized OTOC, denoted as ROTOC, is defined as: $\text{ROTOC} = \text{OTOC}/\text{Echo}_W$. For example, when $\wt{(V)}=1$, the ROTOC is
	\begin{align}
		\frac{\mathbb{E}\left[\widetilde{C}_{W,V}(t)\right]}{\mathbb{E}\left[\braket{\widetilde{W}(t)W(t)}\right]}=\frac{8}{3N}\frac{\sum_w wb_w}{\sum_wb_w}=\frac{8}{3N}\braket{w}_c\ed
	\end{align}
	
	The discussion above shows that once $b_w(t)$ is known, physical observables such as the dressed OTOC and the ROTOC follow straightforwardly. Motivated by this observation, we focus in the remainder of this work on developing a closed dynamical description for $b_w$. To this end, we introduce a generating-function formulation whose evolution encodes the full hierarchy of $b_w$ in the next section, providing a systematic framework for incorporating higher-order corrections in the dilute limit.
	
	\section{Equation for operator size distribution}\label{sec:eq}
	We now turn to the dynamical description of operator growth. As emphasized in the previous section, the central quantity controlling both operator size and dressed OTOC is the distribution $b_w(t)$ of Pauli-string weights.
	
	In this section, we derive the time evolution equation for $b_w$ in the Brownian spin model. Our approach is based on a generating-function formulation, which allows us to organize the dynamics systematically and to incorporate higher-order corrections within the dilute limit. This framework avoids the need to diagonalize large transition matrices and provides direct access to perturbative corrections beyond leading order. Hence it will make transparent both the leading order structure and the mechanisms by which subleading corrections influence late-time behavior.
	
	\subsection{Time evolution of $b_w$}
	We now derive the time evolution equation for the operator weight distribution $b_w(t)$. To this end, we analyze the infinitesimal change of Pauli-string coefficients $c_P(t)$ over a short time interval $\delta t$ under Brownian Hamiltonian evolution, and then perform an ensemble average to obtain a closed dynamical equation for $b_w$.
	
	Expanding the Heisenberg evolution operators to second order in $\delta t$, we obtain
	\begin{align}\label{eq:delta-O}
		\delta O(t)&=e^{iH_{t}\delta t}O(t)e^{-iH_{t}\delta t}-O(t)-\kappa\sum_P \wt_Pc_P(t)P\delta t\nn &=i[H,O(t)]\delta t+\left(HO(t)H-\frac{1}{2}\{H^{2},O(t)\}\right)\delta t^{2}-\kappa\sum_P \wt_Pc_P(t)P\delta t+\mathcal{O}(\delta t^{3})\co
	\end{align}
	so that 
	\begin{align}
		\delta c_P(t)&=\left(P\bigg|i[H,O(t)]\delta t+\left(HO(t)H-\frac{1}{2}\{H^{2},O(t)\}\right)\delta t^{2}\right)-\kappa\sum_P \wt_Pc_P(t)\delta t\;.
	\end{align}
	We use $[A,B]$, $\{A,B\}$, and $(A|B)$ to represent the commutator, anticommutator, and the operator inner product of operators $A$ and $B$, respectively. Recall that our Hamiltonian is given by Eq.~\eqref{eq:sysH}, where the couplings $J_{A}(t)$ are Gaussian and uncorrelated in time, only terms quadratic in the Hamiltonian survive in the ensemble average. As a result, contributions linear in $H$ vanish, and the leading nontrivial evolution arises at $O(\delta t^2)$. Hence
	\begin{equation}\label{eq:deltab_w}
		\begin{aligned} 
			& \mathbb{E}\left(c_{P}(t+\delta t)\widetilde{c}_{P}(t+\delta t)-c_{P}(t)\widetilde{c}_{P}(t)\right)=\mathbb{E}\left(c_{P}\delta\widetilde{c}_{P}+\widetilde{c}_{P}\delta c_{P}+\delta c_{P}\delta\widetilde{c}_{P}\right)\\
			= & \sum_{A,Q}\bigg[\mu_{\mathsf{w}(O_{A})}^{2}\mathbb{E}\left[\left(c_{P}\widetilde{c}_{Q}+\widetilde{c}_{P}c_{Q}\right)\left(P\bigg|\left(O_{A}QO_{A}-\frac{1}{2}\{O_{A}^{2},O(t)\}\right)\right)\right]\delta t\\
			& -r\mu_{\mathsf{w}(O_{A})}^{2}\mathbb{E}\left[c_{Q}\widetilde{c}_{Q}(P|[O_{A},Q])^{2}\right]\delta t\bigg]-2\kappa\sum_{P}\mathsf{w}_{P}\mathbb{E}\left[c_{P}\widetilde{c}_{P}\right]\delta t\ed
		\end{aligned}
	\end{equation}
	Since any two Pauli strings either commute or anticommute, only operators $O_A$ satisfying $\{O_A,Q\}=0$ contribute in Eq.~\eqref{eq:deltab_w}. 
	To ensure the anticommutation relation $\{Q, O_A\} = 0$ between Pauli strings $Q$ and $O_A$, they must differ on an odd number of sites. We therefore analyze how a weight-$n$ Pauli string $O_A$ can be constructed relative to a \textit{fixed} weight-$w$ Pauli string $Q$. Two Pauli operators on a given site anticommute only if they are different and non-identity. An odd number of anticommuting sites is achieved by selecting an odd integer $p$ (with $p \le \min(w, n)$) from the $w$ non-identity sites of $Q$. This choice contributes a combinatorial factor of $\binom{w}{p}$. On each of these $p$ sites, $O_A$ can take either of the two Pauli operators that anticommute with the corresponding operator in $Q$, giving an additional factor of $2^{p}$.
	
	The remaining sites must commute. From the $w-p$ remaining non-identity sites of $Q$, we choose $m$ sites (with $p + m \le n$) where $O_A$ carries the \textit{same} Pauli operator as $Q$, contributing $\binom{w-p}{m}$. The remaining $n - m - p$ Pauli matrices of $O_A$ must then be placed on sites where $Q$ is the identity. We select these $n - m - p$ sites from the $N-w$ identity sites of $V$, which can be done in $\binom{N-w}{n-m-p}$ ways. On each of these chosen sites, $O_A$ can be any of the three non-identity Pauli operators, yielding a factor of $3^{\,n-m-p}$. A schematic illustration of these overlap patterns is provided in Fig.~\ref{fig:overlap}.
	The total number of weight-$n$ Pauli strings that anticommute with a fixed weight-$w$ Pauli string is therefore given by
	\begin{align}\label{eq:C_const}
		C^w_{n}&=\sum_{p\ \text{odd},\ m+p\le n}C_{n,p,m}^{w}\;,\\
		C_{n,p,m}^{w}&\equiv2^{p}\times 3^{n-m-p}\binom{w}{p}\binom{w-p}{m}\binom{N-w}{n-m-p}\;.
	\end{align}
	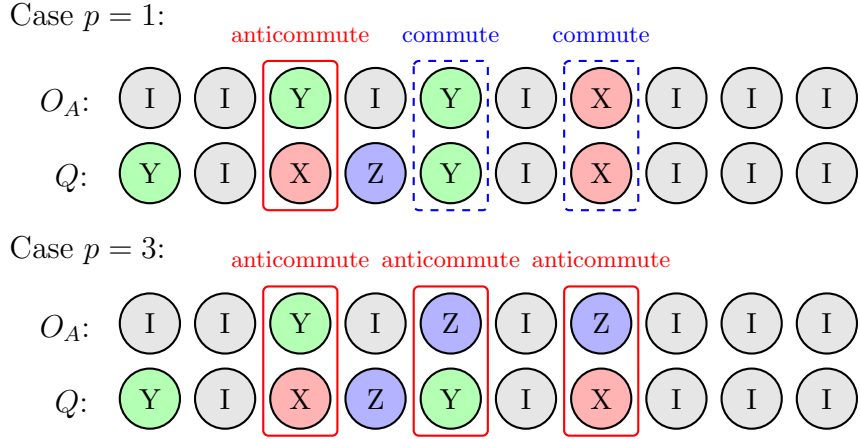
\begin{figure}[ht]
		\centering
		\begin{tikzpicture}[
			site/.style={draw, circle, minimum size=8mm, thick},
			I/.style={fill=gray!20},
			X/.style={fill=red!30},
			Y/.style={fill=green!30},
			Z/.style={fill=blue!30},
			overlap/.style={fill=yellow!50},
			anticommute/.style={draw=red, thick, rounded corners=2pt},
			commute/.style={draw=blue, thick, rounded corners=2pt, dashed},
			label/.style={midway, font=\small}
			]
			
			% ========== CASE p = 1 ==========
			\node[anchor=north west, font=\large, align=center] at (-2, 8.4) {Case $p=1$:};
			
			% O_A: Weight 3
			\node[anchor=north west, font=\large] at (-1.6, 7.25) {$O_A$:};
			\foreach \i/\type in {1/I, 2/I, 3/Y, 4/I, 5/Y, 6/I, 7/X, 8/I, 9/I, 10/I} {
				\node[site, \type] (A\i) at (\i-1, 7) {\type};
			}

			% Q: Weight 5
			\node[anchor=north west, font=\large] at (-1.4, 6.25) {$Q$:};
			\foreach \i/\type in {1/Y, 2/I, 3/X, 4/Z, 5/Y, 6/I, 7/X, 8/I, 9/I, 10/I} {
				\node[site, \type] (Q\i) at (\i-1, 6) {\type};
			}
			
			% Highlight anticommuting site (site 3)
			\draw[anticommute] ($(A3.north west)+(-0.2,0.2)$) rectangle ($(Q3.south east)+(0.2,-0.2)$);
			\node[above=0.2 of A3, font=\footnotesize, red] {anticommute};
			
			% Highlight commuting sites (sites 5 and 7)
			\draw[commute] ($(A5.north west)+(-0.2,0.2)$) rectangle ($(Q5.south east)+(0.2,-0.2)$);
			\node[above=0.2 of A5, font=\footnotesize, blue] {commute};
			
			\draw[commute] ($(A7.north west)+(-0.2,0.2)$) rectangle ($(Q7.south east)+(0.2,-0.2)$);
			\node[above=0.2 of A7, font=\footnotesize, blue] {commute};
			
			% ========== CASE p = 3 ==========
			
			\node[anchor=north west, font=\large, align=center] at (-2, 5.3) {Case $p=3$:};
			
			% O_A: Weight 3
			\foreach \i/\type in {1/I, 2/I, 3/Y, 4/I, 5/Z, 6/I, 7/Z, 8/I, 9/I, 10/I} {
				\node[site, \type] (A2\i) at (\i-1, 4) {\type};
			}
			\node[anchor=north west, font=\large] at (-1.6, 4.25) {$O_A$:};
			
			% Q: Weight 5
			\node[anchor=north west, font=\large] at (-1.4, 3.25) {$Q$:};
			\foreach \i/\type in {1/Y, 2/I, 3/X, 4/Z, 5/Y, 6/I, 7/X, 8/I, 9/I, 10/I} {
				\node[site, \type] (Q2\i) at (\i-1, 3) {\type};
			}
			
			% Highlight all anticommuting sites (sites 3, 5, 7)
			\draw[anticommute] ($(A23.north west)+(-0.2,0.2)$) rectangle ($(Q23.south east)+(0.2,-0.2)$);
			\node[above=0.2 of A23, font=\footnotesize, red] {anticommute};
			
			\draw[anticommute] ($(A25.north west)+(-0.2,0.2)$) rectangle ($(Q25.south east)+(0.2,-0.2)$);
			\node[above=0.2 of A25, font=\footnotesize, red] {anticommute};
			
			\draw[anticommute] ($(A27.north west)+(-0.2,0.2)$) rectangle ($(Q27.south east)+(0.2,-0.2)$);
			\node[above=0.2 of A27, font=\footnotesize, red] {anticommute};
			
		\end{tikzpicture}
		\caption{An illustration of the overlap patterns in case $p=1,m=2$ and case $p=3,m=0$.}
		\label{fig:overlap}
	\end{figure}

	We now evaluate the contributions of these processes to the evolution of $b_w$. The first term on the right-hand side of~\eqref{eq:deltab_w} yields
	\begin{equation}
		\begin{aligned}
			{d\over dt}b_w\supset &\sum_{A,P,Q|\{O_A,Q\}=0,\wt(P)=w}-4\mu_{\mathsf{w}(O_{A})}^{2}\mathbb{E}\left[c_P(t)\widetilde{c}_Q(t)(P|QO_A^2)\right]=-4\sum_n\mu_n^2C^w_nb_w\ed
		\end{aligned}
	\end{equation}
	The contribution of the second term of~\eqref{eq:deltab_w} is
	\begin{equation}
		\begin{aligned}
			\frac{d}{dt}b_{w}\supset&\sum_{A,P,Q|\{O_{A},Q\}=0,\wt(P)=w}-r\mu_{\mathsf{w}(O_{A})}^{2}\mathbb{E}\left[c_{Q}\widetilde{c}_{Q}(P|2QO_{A})^{2}\right]\\&=\sum_{n}\sum_{A,Q|\{O_{A},Q\}=0,\wt(O_{A})=n,\wt(QO_{A})=w}4\mu_{n}^{2}rc_{Q}\widetilde{c}_{Q}\\&=\sum_{n}4\mu_{n}^{2}r\left(\sum_{p\text{ odd},\ m+p\le n}C_{n,p,m}^{w+2m+p-n}b_{w+2m+p-n}\right)
		\end{aligned}
	\end{equation}
	where we have used the production of $Q$ and $O_A$ is $\pm i^p$ times a weight $w+n-2m-p$ Pauli string. 
	The last term in Eq.~\eqref{eq:deltab_w} just gives a decay term
	\begin{align}
		\frac{d}{dt}b_{w}\supset - 2\kappa w b_w\ed
	\end{align}
	Finally, we obtain
	\begin{align}\label{eq:dbw}
		\frac{db_{w}}{dt}=\sum_{n=2}^L \,4\mu_{n}^{2}\left[\sum_{p\ \text{odd},p+m\le n} r C_{n,p,m}^{w+2m+p-n} \,b_{w+2m+p-n}-C_{n}^{w}b_{w}\right]-2\kappa  w b_w\ed
	\end{align}
	The evolution can be written in matrix form as $\frac{db_w}{dt} = \sum_{w'=1}^{N} M_{ww'} b_{w'}$. Direct diagonalization of the $N \times N$ matrix $M$ is generally intractable. Following the approach outlined in the introduction, we circumvent this difficulty by formally extending $M$ to an infinite-dimensional matrix $M_\infty$, while retaining explicit $N$-dependence in its entries. Correspondingly, the initial distribution is extended to infinite dimensions by setting $b_{w>N}(0)=0$. We then define an approximate generating function
	\begin{align}
		\mathsf{G}(x,t)=X_\infty^T.e^{M_\infty t}\boldsymbol{b}_\infty(0),~X_{\infty}\equiv [x,x^2,\ldots]\ed
	\end{align}   
	Replacing each power $w^k$ by the operator $(x\partial_x)^k$ in Eq.~\eqref{eq:dbw} leads to the partial differential equation for $\mathsf{G}(x,t)$
	\begin{align}
		\partial_t \mathsf{G}(x,t)=\mathcal{M}_x \mathsf{G}(x,t)\ed
	\end{align}
	We then expand all quantities in powers of $1/N$; for example, the generating function is written as $\mathsf{G}(x,t)=\sum_{k=0}^\infty \mathsf{G}^{(k)}(x,t) N^{-k}$. The eigenvalue problem is subsequently solved order by order using perturbation theory:
	\begin{align}
		\mathcal{M}_x G_k(x,t)=\lambda_k G_k(x,t)\ed
	\end{align}
	At leading order in $1/N$ within the dilute limit, the evolution matrix becomes strictly lower triangular, enabling an explicit spectral solution. We now analyze this limit and solve the dynamics using the generating-function method.
	
	\subsection{Leading order solution}\label{sec:solutiondilutelimit}
	We now specialize to the dilute limit, where the operator-weight distribution is concentrated at weights $w \ll N$. In this regime, the  factor $C^{w}_{n,p,m}$ scales as $\mathcal{O}(N^{n-m-p})$, so the dominant contributions come from terms with the smallest possible $p$ and $m$. Because anticommutation requires $p$ to be odd, the leading contribution corresponds to $p=1$ and $m=0$. Retaining only this $C^{w}_{n,1,0}$ term gives
	\begin{align}
		\frac{d b_w}{d t}=4\sum_{n=2}^{L}\mu_n^2[rC_{n,1,0}^{w-(n-1)}b_{w-(n-1)}-C^w_{n,1,0}b_w]-2\kappa  w b_w+\mathcal{O}(1/N)\;.
	\end{align}
	In the large-$N$ limit, we scale the couplings as $\mu_n^2 = \frac{a_{n}(n-1)!}{4 \times 3^{\,n-1} N^{\,n-1}}$ so that the leading-order equation in the dilute limit becomes
	\begin{align}\label{eq:dbw_LO}
		\frac{db_{w}}{dt} = -2w\bigl(a_\Sigma + \kappa\bigr) b_{w}
		+ 2r \sum_{n=2}^{L} \bigl[w - (n-1)\bigr] a_{n} b_{w-(n-1)} + \mathcal{O}(1/N)
	\end{align}
	where $a_\Sigma = \sum_{n=2}^L a_n$. Writing the evolution in matrix form
	\begin{align}
		\frac{db_{w}}{dt} = \sum_{w'=1}^N M^{(0)}_{ww'} b_{w'} + \mathcal{O}(1/N)\co
	\end{align}
	we find that the resulting generator $M^{(0)}$ is strictly lower triangular in the weight basis. Consequently, its spectrum is given directly by the diagonal entries, yielding eigenvalues
	\begin{equation}
		\lambda_l^{(0)} = -2l\,(a_\Sigma + \kappa), \qquad l = 1,2,3,\dots .
	\end{equation}
	
	To facilitate the generating-function analysis, we extend the evolution matrix $M^{(0)}$ to infinite dimensions by formally letting the weight index $w \to \infty$, obtaining $M_\infty^{(0)}$. Denoting the $k$-th eigenvector of this extended matrix by ${v}_k$, we define its generating function as
	\begin{align}
		G_k^{(0)}(x) \equiv \sum_{j=1}^{\infty} v_{k;j}\,x^{j}
	\end{align}
	where $v_{k;j}$ denotes the $j$-th component of ${v}_k$. The equation for $G^{(0)}_k(x)$ follows from the eigenvector equation for ${v}_k$,
	\begin{equation}
		-(a_\Sigma+\kappa)x\partial_x G_k^{(0)}(x) + r\sum_{n=2}^L a_n x^n \partial_x G_k^{(0)}(x) = -k(a_\Sigma+\kappa)G_k^{(0)}(x)\ed
	\end{equation}
	Its solution is
	\begin{align}
		G_{k}^{(0)}(x) = A_k \exp\!\left(\int_{1}^{x} ds\, \frac{(a_\Sigma+\kappa)k}{(a_\Sigma+\kappa)s - r\sum_{n=2}^{L} a_n s^{n}}\right) = \bigl(G_1^{(0)}(x)\bigr)^k\ed
	\end{align}
	Analogous to an eigenvector decomposition, the generating function of the initial distribution can be expanded in the eigenbasis $\{G_k^{(0)}(x)\}$ as
	\begin{align}
		\mathsf{G}_{\mathrm{init}}(x) \equiv \sum_{k=1}^{\infty} b_k(0) x^k = \sum_{j=1}^{\infty} c_j G_j^{(0)}(x)\ed
	\end{align}
	Consequently, the time-evolved generating function at leading order is
	\begin{align}
		\mathsf{G}^{(0)}(x,t) = \sum_{k=1}^{\infty} b_k(t) x^k = \sum_{j=1}^\infty e^{-2j(a_\Sigma+\kappa)t} c_j G_j^{(0)}(x)\ed
	\end{align}
	The expansion coefficients $c_l$ are determined via the generating function of the left eigenvectors. Because $M^{(0)}_\infty$ is non-Hermitian, its left and right eigenvectors are distinct. Denoting the $k$-th left eigenvector by ${w}_k$, it satisfies
	\begin{align}
		\sum_i w_{k;i} M_{ij}^{(0)} = \lambda_k w_{k;j}\ed
	\end{align}
	Since $M_\infty^{(0)}$ is lower triangular, ${w}_k$ has non-zero entries only for indices $i \le k$. Proceeding as before, we define the left-eigenvector generating function
	\begin{align}
		W_k^{(0)}(x) = \sum_{j=1}^\infty w_{k;j} x^j\ed
	\end{align}
	Its governing equation reads
	\begin{align}
		-x(a_\Sigma+\kappa)\frac{d}{dx}W^{(0)}_k
		+ r\sum_{n=2}^L a_n x \frac{d}{dx}\!\left(\frac{W_k^{(0)}}{x^{\,n-1}} - \frac{W_k^{{(0)}\le n-2}}{x^{\,n-1}}\right)
		= k(a_\Sigma+\kappa) W_k^{(0)}
	\end{align}
	where $W_k^{{(0)}\le m}(x) = \sum_{j=1}^{m} w_{k;j} x^j$. The function $W_k^{(0)}(x)$ is typically a polynomial of finite degree.
	As an alternative to solving the differential equation directly, $W_k^{(0)}(x)$ can be constructed recursively from the coefficients $w_{k;j}$.
	The bi-orthogonality condition can then be imposed as
	\begin{align}
		\braket{W_j^{(0)}|G_k^{(0)}} 
		= \frac{1}{2\pi i}\oint_{z=0}\frac{\dd z}{z}\,W^{(0)*}_j(1/z)\,G_k^{(0)}(z)=\delta_{jk}
	\end{align}
	where $W^{(0)*}_j(x)$ denotes complex conjugation of every parameter in the function except the formal variable $x$. Here the bracket on the left-hand side denotes a bi-orthogonal pairing between left and right generating functions, defined via a contour integral. This biorthogonal normalization ensures a complete spectral decomposition despite the non-Hermitian nature of $M^{(0)}_\infty$.
	In principle, the expansion coefficients $c_l$ can be obtained from
	\begin{align}
		c_j = \braket{W_j^{(0)}|G_{\text{init}}}\ed
	\end{align}
	However, a more direct approach is to introduce a time-dependent variable $x_t(x,t)$ defined implicitly by
	\begin{align}
		G_1^{(0)}(x_t) = G_1^{(0)}(x)\,e^{-2(a_\Sigma+\kappa)t} \ed
		\label{eq:xt}
	\end{align}
	With this definition, the time evolution of the generating function takes the simple form
	\begin{align}
		\mathsf{G}^{(0)}(x,t)=\mathsf{G}_{\text{init}}(x_t)\ed 
	\end{align}
	This framework holds for any initial size distribution, showing that the generating function approach provides a powerful method for deriving analytical results for arbitrary initial conditions and interaction types. The main challenge lies in obtaining an explicit expression for \( x_t \) by solving Eq.~\eqref{eq:xt}. For many scenarios, such as systems with mixed two body and three body interactions, an exact analytical expression for \( x_t \) may not be available. Nevertheless, Eq.~\eqref{eq:xt} can always be treated as a definition of \( x_t \), which remains useful for numerical computations.
	
	Once the generating function \( \mathsf{G}^{(0)}(x,t) \) is determined, physical observables follow directly. For example, \( \langle w \rangle_c \) used in the calculation of OTOC can be easily obtained as
	\begin{align}
		\langle w \rangle_c=\frac{\sum_{j}wb_w(t)x^w|_{x=1}}{\sum_{w}b_w(t)}=\frac{\partial_x \mathsf{G}^{(0)}(x,t)}{\mathsf{G}^{(0)}(x,t)}\bigg|_{x=1}+\mathcal{O}(1/N) \;.
	\end{align}

	\subsection{Two-body and three-body interactions}\label{sec:23-body}
	In this subsection, we analyze the dilute-limit solution for systems with two-body and three-body interactions. 
	\paragraph{Two-body interactions} We begin with the two-body case, which has been studied previously in~\cite{Nadie2025}, and use it as a benchmark to validate our formalism and clarify the structure of the generating-function approach. For $L=2$, the eigenvalue equation for the generating function reduces to
	\begin{align}
		-2(a_2+\kappa)x\partial_xG^{(0)}_k(x)+2ra_2 x^2\partial_xG_k^{(0)}(x)=\lambda_k^{(0)}G_k^{(0)}(x)\;.
	\end{align}
	Requiring the solution to admit a power-series expansion in positive integer powers of $x$ enforces the quantization condition
	\begin{equation}
		\lambda_k^{(0)}=-2k(a_2+\kappa) \;,\; k \in \mathbb{Z}^{+}\;,
	\end{equation}
	in agreement with the general triangular structure identified in~Section \ref{sec:solutiondilutelimit}. We fix the normalization by setting the coefficient off $x^k$ in the expansion of the $k$-th eigenfunction to unity. The resulting eigenfunctions take the simple form
	\begin{align}
		G_k^{(0)}(x)=\frac{x^k}{(1-r_\text{eff}x)^k}\;,\; \ r_{\text{eff}}\equiv\frac{a_2 r}{a_2+\kappa}\;.
	\end{align}
	Notice that the generating functions factorize as 
	\begin{equation}
		G^{(0)}_k(x)=(G_1^{(0)}(x))^k\;,
	\end{equation}
	a property that greatly simplifies the time evolution and, as we show below, extends to more general interaction structures. 
	The time-dependent generating function becomes
	\begin{align}
		\mathsf{G}(x,t)=\sum_{l}c_lG_1(x_t)^l=G_{\text{init}}(x_t)\;,x_t=\frac{xe^{-2(a_2+\kappa)t}}{1-r_{\mathsf{eff}}x(1-e^{-2(a_2+\kappa)t})}\;.
	\end{align}
	which is valid for arbitrary initial distributions.
	For an operator with an initial distribution $b_w(0)=\delta_{w,w_0}$, we have $\mathsf{G}(x,t)=x^{w_0}_t$. Hence
	\begin{align}
		\braket{w}_c=\left.\frac{\partial_x \mathsf{G}(x,t)}{\mathsf{G}(x,t)}\right|_{x=1}=\frac{w_0}{1-r_{\text{eff}}+r_{\text{eff}}e^{-2(a_2+\kappa)t}}\;,
	\end{align}
	At late times, 
	\begin{align}\label{eq:two-body-large-t}
		\braket{w}_c|_{t\to\infty}=w_0/(1-r_{\text{eff}})\;.
	\end{align}
	This solution reproduces the known leading-order behavior in \cite{Nadie2025}.

	\paragraph{Three-body interactions} We now turn to systems with purely three-body interactions. Unlike the two-body case, three-body interactions induce operator weight changes in steps of $\pm 2$, leading to qualitatively new structural features in the dynamics even at leading order.
	The generating function for the $k$-th eigenvector is found to be
	\begin{align}
		G_{k}^{(0)}(x)=\left(\frac{x}{\sqrt{1-r_{\text{eff}}x^{2}}}\right)^{k}\co
	\end{align}
	where we have introduced the effective imperfection parameter $r_\text{eff} = \frac{a_3r}{a_3+\kappa}$. 
	Correspondingly, the flow of the generating-function argument is given by
	\begin{align}
		x_t(x,t)=\frac{xe^{-2t(a_3+\kappa)}}{\sqrt{1+r_\text{eff}\,x^2(e^{-4t(a_3+\kappa)}-1)}}\ed
	\end{align}
	We note that the expansion of \( x_t \) in \( x \) contains only odd powers of \( x \), which reflects the fact that under three-body interactions, the change in operator weight \( b_w \) occurs in steps of \( \pm 2 \). More specifically, considering an initial operator with weight \( w_0 \), its generating function is simply given by \( x_t^{w_0} \). Expanding this reveals that the powers of \( x \) share the same parity as \( w_0 \). The leading order calculation gives the late-time behavior
	\begin{align}
		\langle w \rangle_c\vert_{t\to\infty} = w_0/(1-r_{\text{eff}})\ed 
	\end{align}
	This result is consistent with that obtained in two-body interactions.

	\section{Higher-order effects: Two-body interactions}\label{sec:L2}
	As noted in the previous section, the evolution matrix governing the operator weight distribution is lower triangular at leading order in $N$. As a result, the $k$-th eigenvector has support only on components $v_{k,j}$ with $j\ge k$, and its eigenvalue is given by $\lambda_k^{(0)} = -2k(\kappa+a)$. Consequently, the late-time dynamics predicted at leading order is controlled solely by the smallest nonzero eigenvalue present in the initial state.
	
	However, this prediction is generically violated in numerical simulations at finite $N$. Although $1/N$ corrections are parametrically small, they qualitatively alter the late-time dynamics by enabling population transfer between eigenmodes that are decoupled at leading order. At sufficiently long times, these higher-order effects dominate and determine the true asymptotic behavior. Accurately capturing the late-time operator growth therefore requires a systematic inclusion of $1/N$ corrections.
	
	\subsection{First order perturbation}
	We now treat the $1/N$ terms as a perturbation to the leading order generator and compute the resulting corrections to both the eigenvalues and eigenfunctions. The full evolution equation of $b_w$ turns to be
	\begin{equation}\label{eq:bw-eq-two-body}
		\begin{aligned}
			\frac{db_{w}}{dt}& = -\frac{2w\left((w-1)+3(N-w)\right)}{3N}b_{w}-2w\kappa b_{w}\\
			&\newline + r\left[\frac{2(N-w+1)(w-1)}{N}b_{w-1}+\frac{2w(w+1)}{3N}b_{w+1}\right].
		\end{aligned}
	\end{equation}
	For notational simplicity, we set $a_2=1$ in the following. The general case can be recovered by replacement $t\to a_2 t$ and $\kappa\to \kappa/a_2 $. We now recast this equation in terms of the generating function $\mathsf{G}(x,t)$
	\begin{align}
		\partial_t \mathsf{G}(x,t)=&\left(2rx^{2}\partial_{x}-2(\kappa+1)x\partial_{x}\right)\mathsf{G}(x,t) \notag \\
		&+\frac{2}{3N}\left(-3r\left(x^{3}\partial_{x}^{2}+x^{2}\partial_{x}\right)+rx\partial_{x}^{2}+2x^{2}\partial_{x}^{2}+3x\partial_{x} \right)\mathsf{G}(x,t)\;.
	\end{align}
	We denote it as
	\begin{align}
		\partial_t \mathsf{G}(x,t)=\left[\widehat{A}_0 + {\widehat{A}_1\over N}\right] \mathsf{G}(x,t)\;.
	\end{align}
	We now employ perturbation theory to compute the corrections to the eigenvector generating function and the eigenvalues. Let $G_k^{(0)}(x)$ and $\lambda_k^{(0)}$ denote the results obtained in the previous chapter. Let $G_k^{(n)}(x)$ and $\lambda_k^{(n)}$ denote the $n$-th order corrections, The first-order corrections are given by:
	\begin{align}
		\widehat{A}_{0}G_k^{(1)}(x)+\widehat{A}_{1}G_k^{(0)}(x)=\lambda_k^{(1)}G_k^{(0)}(x)+\lambda_k^{(0)}G_k^{(1)}(x)\ed 
	\end{align}
	Solving the equation and imposing the condition $G_{k}^{(1)}(0)=0$, we have
	\begin{equation}
		\begin{aligned}
			\lambda_k^{(1)}&=\frac{2 k \left(\kappa +k \left(2 \kappa +3
				r^2+2\right)-r^2+1\right)}{3 (\kappa +1)}\equiv B^{(1)}_{2}k^2+B^{(1)}_{1}k \;,\\
			G_k^{(1)}(x)&=-\frac{k(1+\kappa)^kx^{k-1}}{6(\kappa+1)(\kappa-rx+1)^{k+2}}\bigg[2(3k-1)r^{3}x^{2}-3(\kappa+1)(3k-1)r^{2}x\\
			&\newline+2(\kappa+1)r\left(-\kappa+k\left(\kappa+2x^{2}+1\right)+(3\kappa+4)x^{2}-1\right)\\
			&\newline+(\kappa+1)^{2}x(-3\kappa+3(\kappa-1)k-7)\bigg]+C_k^{(1)} G_k^{(0)}(x)
		\end{aligned}
	\end{equation}
	where the constant $C_k^{(1)}$ reflects a freedom in the normalization of eigenvectors and does not affect physical observables. Similar to the zeroth-order calculation, self-consistency yields the result for $\lambda_k^{(1)}$. We can impose $\braket{W_k^{(0)}|G_k^{(1)}}=0$ to determine $C_k^{(1)}$. It is easy to find 
	\begin{align}
		W_k^{(0)}(x)= x^k(1-r_{\text{eff}}/x)^{k-1}\ed 
	\end{align}
	Direct calculation gives
	\begin{align}
		C_{k}^{(1)}=-\frac{k\left(3\kappa^{2}+10\kappa+3k\left(-\kappa^{2}+r^{2}+1\right)+3r^{2}+7\right)}{6(\kappa+1)^{2}}\equiv R_1^{(1)} k +R_2^{(1)} k^2\;,
	\end{align}
	where
	\begin{align}
		R_{1}^{(1)}=-\frac{3 \kappa ^2+10 \kappa +3 r^2+7}{6(\kappa+1)^{2}}\;,\;R_{2}^{(1)}=-\frac{-\kappa^{2}+r^{2}+1}{2(\kappa+1)^{2}} \;.
	\end{align}
	Now we want to find an operator $\widehat{O}_x^{(1)}$, such that $\widehat{O}_x^{(1)}G_k^{(0)}(x)=G_k^{(1)}(x)$. First, we express $G_k^{(1)}(x)$ in the form $G_k^{(1)}(x)=f_1(x)kG_k^{(0)}(x)+f_2(x)k^2G_k^{(0)}(x)$. By substituting $kG^{(0)}_k(x)=\widehat{A}_0 G^{(0)}_k(x)/\lambda_1^{(0)}$, one can finally find
	\begin{align}
		G_k^{(1)}(x)=\left(f(x)\partial_x+g(x)\partial_x^2+R_2^{(1)}\frac{\widehat{A}^2_0}{\left(\lambda_1^{(0)}\right)^2}+R_1^{(1)}\frac{\widehat{A}_0}{\lambda_1^{(0)}}\right)G_k^{(0)}(x)\equiv \widehat{O}_x^{(1)} G_k^{(0)}(x)\;,
	\end{align}
	with 
	\begin{align}
		f(x)=h_1x+h_2x^2 +h_3 x^3,g(x)=s_1 x + s_2 x^2 + s_3 x^3+s_4x^4\;,
	\end{align}
	\begin{align}
		&h_{1}=\frac{2 r^2}{3 (\kappa +1)^2},h_{2}=-\frac{(3 \kappa -4) r }{3 (\kappa +1)^2},h_{3}=-\frac{r\left(3 r^3-3 (\kappa -1) (\kappa +1) r\right)}{3 (\kappa +1)^4}\;,\nn
		&s_{1}=-\frac{r}{3 \kappa +3},s_{2}=\frac{r^2}{(\kappa +1)^2},s_{3}=-\frac{(3 \kappa -1) r}{3 (\kappa +1)^2},s_4=-\frac{r^2 \left(-\kappa ^2+r^2+1\right)}{2 (\kappa +1)^4}\;.
	\end{align}
	By applying the same approach, we can derive an operator $\widehat{\Lambda}_x^{(1)}$ such that
	\begin{align}
		\widehat{\Lambda}_x^{(1)}G_k^{0}(x)=\lambda_k^{(1)}G_k^{(0)}(x)\;.
	\end{align}
	The expansion coefficients is calculated by $c_l =\braket{W_l|\mathsf{G}_{\text{init}}}$. 
	We now expand $\mathsf{G}_{\text{init}}(x)$ to first order in $1/N$ and we have
	\begin{align}\label{eq:Ginit-exp}
		\sum_l\left(c_l^{(0)}G^{(0)}_l(x)+\frac{1}{N}\left(c_l^{(1)}G_l^{(0)}(x)+c_l^{(0)}G_l^{(1)}(x)\right)\right)=\mathsf{G}_{\text{init}}(x)\;.
	\end{align}
	Since $\{W_i^{(0)},G_i^{(0)}(x)\}$ forms a complete basis, we have $\sum_l c_l^{(0)}G^{(0)}_l(x)=\mathsf{G}_{\text{init}}(x)$. Then using $ G^{(1)}_k(x)=\widehat{O}_x^{(1)}G_k^{(0)}(x)$, we have
	\begin{align}
		\sum_{l}c_{l}^{(1)}G_{l}^{(0)}(x)+\widehat{O}_{x}^{(1)}\mathsf{G}_{\text{init}}(x)=0\;.
	\end{align}
	Then, replace $x\to x_t$ leads to
	\begin{align}\label{eq:eqNL}
		\sum_le^{\lambda_l^{(0)}t}c_l^{(1)}G_l^{(0)}(x)+\left(\widehat{O}_x^{(1)}\mathsf{G}_{\text{init}}(x)\big|_{x\to x_t}\right)=0 \;.
	\end{align}
	We now expand $\mathsf{G}(x,t)$ to first order in $1/N$
	\begin{equation}\label{eq:two-body_G1}
		\begin{aligned}
			\mathsf{G}(x,t)&=\sum_{k=1}^{\infty}c_{k}e^{\left(\lambda_{k}^{(0)}+\frac{\lambda_{k}^{(1)}}{N}\right)t}\left(G_k^{(0)}(x)+\frac{G_k^{(1)}(x)}{N}\right)+ \mathcal{O}(N^{-2})
			\\&=\mathsf{G}_{\text{init}}(x_t)+\frac{1}{N}\sum_{k=1}^\infty\left( c_k^{(0)}\lambda_k^{(1)}t e^{\lambda_k^{(0)}t}G_k^{(0)}(x)\right. \\
			&\quad \left.+
			\widehat{O}_x^{(1)}c_k^{(0)}e^{\lambda_k^{(0)}t}G_k^{(0)}(x)+c_k^{(1)}e^{\lambda_k^{(0)}t}G_k^{(0)}(x) \right)+ \mathcal{O}(N^{-2})\\
			&=\mathsf{G}_{\text{init}}(x_t)+\frac{1}{N}\left(t\widehat{\Lambda}^{(1)}+\widehat{O}_x^{(1)}\right)\mathsf{G}_{\text{init}}(x_t)-\frac{1}{N}\widehat{O}_x^{(1)}\mathsf{G}_{\text{init}}(x)|_{x\to x_t}+ \mathcal{O}(N^{-2}) \;.
		\end{aligned}
	\end{equation}
	where we have used Eq.~\eqref{eq:eqNL} to obtain the last line.  So that the first order correction is
	\begin{align}\label{eq:gen_NL}
		\mathsf{G}^{(1)}(x,t)=\frac{1}{N}\left(t\widehat{\Lambda}^{(1)}+\widehat{O}_x^{(1)}\right)\mathsf{G}_{\text{init}}(x_t)-\frac{1}{N}\widehat{O}_x^{(1)}\mathsf{G}_{\text{init}}(x)|_{x\to x_t}\ed
	\end{align}
	One can verify that the effects of $C_k^{(1)}$ do not affect the final result. For its contribution to operator $\what{O}_x^{(1)}$ is 
	\begin{align}
		R_2^{(1)}\frac{\widehat{A}^2_0}{(\lambda_1^{(0)})^2}+R_1^{(1)}\frac{\widehat{A}_0}{\lambda_1^{(0)}}\subset \what{O}_x^{(1)}
	\end{align}
	Notice that 
	\begin{align}
		\what{A}_0^k \mathsf{G}_{\text{init}}(x_t)= \what{A}_0^k \mathsf{G}_{\text{init}}(x)\vert_{x\to x_t}\co
	\end{align}
	so they are canceled in Eq.~\eqref{eq:two-body_G1}. 
	For an initial distribution $b_w=\delta_{w,2}$, the first-order perturbative expansion yields superior agreement with the numerical results in the late-time regime, as evidenced by Fig.~\ref{fig:rho}.
	
	\subsection{Second order perturbation}
	Using perturbation theory, we obtain the expression for $G_k^{(2)}$ and $\lambda_k^{(2)}$ with explicit expression given in Appendix \ref{appdix}.  
	Now we expand $\mathsf{G}(x,t)$ to the second order in $1/N$ to obtain
	\begin{align}
		\mathsf{G}^{(2)}(x,t)=&\sum_{k=1}^{\infty}\bigg[\bigg(c_{k}^{(0)}\lambda_{k}^{(2)}t+c_{k}^{(0)}\left(\lambda_{k}^{(1)}\right)^{2}\frac{t^{2}}{2}+c_{k}^{(1)}\lambda_{k}^{(1)}t+c_{k}^{(2)}\bigg)G_k(x)\nn
		&\newline+\left(c_{k}^{(0)}\lambda_{k}^{(1)}t+c_{k}^{(1)}\right)G^{(1)}_k(x)+c_{k}^{(0)}G_k^{(2)}(x)\bigg]e^{\lambda_k^{(0)} t}\;.
	\end{align}
	As before, we replace $kG^{(0)}_k(x)$ with the operator $\widehat{A}_0/\lambda_1^{(0)}G^{(0)}_k(x)$, and we let $\Lambda_x^{(i)},\ \widehat{O}_x^{(i)}$ denote the operator corresponding to $\lambda_k^{(i)},\ G^{(i)}_k(x)$.
	\begin{equation}
		\begin{aligned}
			\mathsf{G}^{(2)}(x,t)=&\left(t\Lambda^{(2)}_x+\frac{t^2}{2}\left(\Lambda_x^{(1)}\right)^2 +\widehat{O}_x^{(2)}+t\widehat{O}_x^{(1)}\Lambda_x^{(1)}  \right)\mathsf{G}_{\text{init}}(x_t) \\
			&+\left(t\Lambda_x^{(1)}+\widehat{O}_x^{(1)}\right)\sum_k c_k^{(1)}e^{\lambda_k^{(0)}t}G_k^{(0)}(x)  
			+\sum _kc_k^{(2)}e^{\lambda_k^{(0)}t}G_k^{(0)}(x)\;.
		\end{aligned}
	\end{equation}
	Like what we do in Eq.~\eqref{eq:Ginit-exp}, expanding $\mathsf{G}_{\text{init}}(x)$ to the second order leads to
	\begin{align}
		\sum_k\left(c_k^{(2)}+c_k^{(1)}\widehat{O}^{(1)}_x+c_k^{(0)}\widehat{O}_x^{(2)}   \right)G^{(0)}_k(x)=0 \;.
	\end{align}
	We then replace $x$ with $x_t$, including those within differential operators
	\begin{align}              
		\sum_k\left(c_k^{(2)}e^{\lambda_k^{(0)}t} G_k^{(0)}(x)+c_k^{(1)}\widehat{O}^{(1)}_{x_t}G_k^{(0)}(x_t)+c_k^{(0)}\widehat{O}_{x_t}^{(2)}G_k^{(0)}(x_t)   \right)&=0 \;.
	\end{align}
	We can now convert the summations over $c_k^{(2)}$ and $c_k^{(1)}$ into summations over $c_k^{(0)}$, thus obtaining the $1/N^2$ correction term as follows
	\begin{equation}\label{eq:gen_NNL}
		\begin{aligned}
			\mathsf{G}^{(2)}(x,t)=&\left(t\Lambda^{(2)}_x+\frac{t^2}{2}\left(\Lambda_x^{(1)}\right)^2 +\widehat{O}_x^{(2)}+\widehat{O}_x^{(1)}\Lambda_x^{(1)}  \right)\mathsf{G}_{\text{init}}(x_t) \\
			&-\left(t\Lambda_x^{(1)}+\widehat{O}_x^{(1)}\right)\left(\widehat{O}^{(1)}_x\mathsf{G}_{\text{init}}(x)\big|_{x\to x_t}\right)-\widehat{O}_x^{(2)}\mathsf{G}_{\text{init}}(x)\big|_{x\to x_t} \\
			&+\left(\widehat{O}_x^{(1)}\widehat{O}^{(1)}_x\mathsf{G}_{\text{init}}(x)\right)\big|_{x\to x_t} \;.
		\end{aligned}
	\end{equation}
	\subsection{Numerical simulation and large-time behavior}
\begin{figure}[ht]
		\begin{center}
			\includegraphics[width=0.4\textwidth]{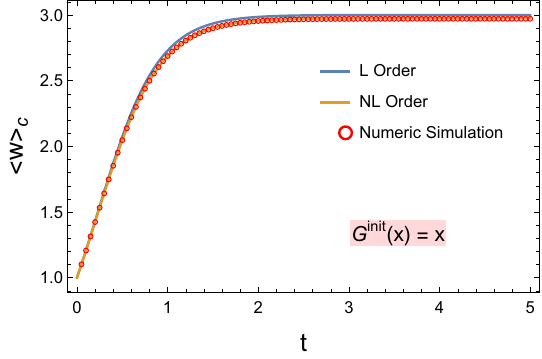}
			\includegraphics[width=0.394\textwidth]{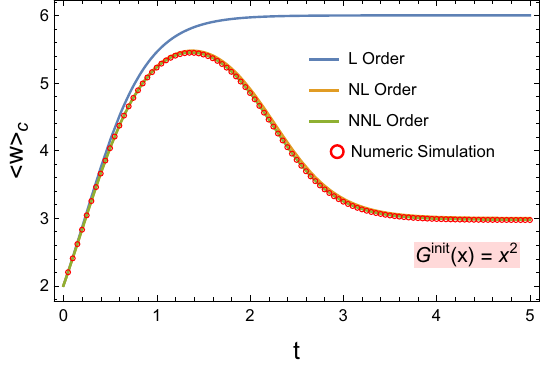}
			\includegraphics[width=0.4\textwidth]{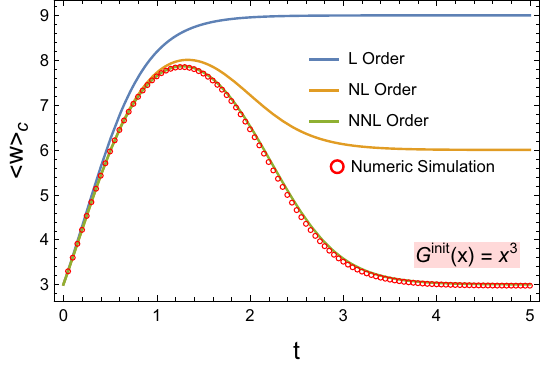}
			\includegraphics[width=0.4\textwidth]{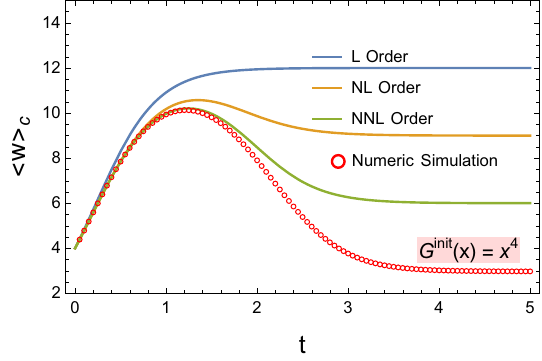}
			\caption{Perturbative predictions versus numerical simulations for different initial operator weights $w_0$. Parameters: $N=100$, $\kappa=0.5$, $r=1$, $a_2=1$, and $a_{j>2}=0$. All four cases converge to the same universal late-time plateau. }
			\label{fig:rho}
		\end{center}
	\end{figure}

	To assess the necessity of higher-order corrections, we compare perturbative predictions for different initial operator weights with numerical simulations for Eq.~\eqref{eq:bw-eq-two-body}. Fig.~\ref{fig:rho} shows the time evolution of the average operator weight $\braket{w}_c$ for several initial conditions.
	
	We observe that the agreement between theory and numerics depends sensitively on both the initial operator weight and the perturbative order retained. For an initial operator of weight $w=1$, the zeroth-order  prediction already captures the late-time behavior accurately. For $w=2$, however, the zeroth-order result deviates substantially at late times, while inclusion of first- and second-order corrections restores agreement. 
	For $w=3$, the second-order calculation remains consistent with numerical simulations. For $w=4$, however, the second-order result deviates significantly beyond the short-time regime. This demonstrates that higher-order corrections become essential for accurately describing the growth of larger initial operators.
	
	This systematic pattern indicates that while the dilute-limit approximation correctly captures the dynamics of initially small operators, it fails in a controlled and hierarchical manner as the initial operator weight increases. Higher-order corrections are required not because the perturbative expansion breaks down, but because the leading order dynamics artificially decouples sectors that are weakly coupled at finite $N$.

	The leading-order (LO) solution predicts a steady-state plateau (Eq.~\eqref{eq:two-body-large-t}) whose magnitude exceeds the initial operator size. This can be understood from the structure of the LO evolution equation (Eq.~\eqref{eq:dbw_LO}), whose right-hand side contains only terms that decrease the weight -- effectively describing the propagation of probability toward larger weights (i.e., to the right). In the full dynamical equation, however, terms corresponding to propagation toward smaller weights also appear but are suppressed by factors of \( 1/N \). Consequently, leftward propagation enters only through higher-order corrections. At asymptotically long times (\( t \to \infty \)), these subleading contributions become significant, implying that the LO result generally fails to capture the correct late-time behavior, even in the dilute limit. Accurate late-time dynamics can only be recovered by including these higher-order terms. 
	
	The phenomenon shown in the figure can be understood within the interaction picture: the $n$-th order perturbative correction essentially corresponds to inserting $n$ interaction vertices $M^{I}_\infty$ at arbitrary positions into the zeroth-order time-evolution operator ($\exp(M_\infty^{(0)}t)$) and integrating over their insertion times:	
	\begin{align}\label{eq:U-exp}
		(e^{M_\infty t})^{(n)}=\mathcal{T}\int_{0}^{t}d^{n}se^{M_{\infty}^{(0)}(t-s_{n})}M_{\infty}^{I}e^{M_{\infty}^{(0)}(s_{n}-s_{n-1})}\cdots M_{\infty}^{I}e^{M_{\infty}^{(0)}s_{1}}
	\end{align}
	where $\mathcal{T}$ means to take time order. Each insertion of $M_\infty^{I}$ shifts the operator-size distribution one step to the left, and each perturbative correction is suppressed by a factor of $(t/N)^n$. Therefore, in the short-time regime, these effects are negligible.
	
	However, when $t \gg N$, i.e., in the late-time limit, higher-order terms in the perturbative expansion become dominant. As a result, regardless of the initial size distribution, $M_\infty^I$ repeatedly shifts weight toward smaller sizes, eventually concentrating it near $w = 1$. Afterward, under prolonged zeroth-order evolution $e^{M_\infty^{(0)}t}$, then $\langle w \rangle_c$ approaches a steady-state plateau at $1/(r-r_{\text{eff}})+\mathcal{O}(1/N)$.
	
	This also explains why accurately capturing the late-time operator growth for an initial operator of weight $m$ necessitates perturbative corrections up to order $m-1$: precisely $m-1$ leftward shifts are required to bring the distribution to its minimal weight. This clarifies the origin of the delayed approach to the asymptotic plateau observed in numerical simulations.
	
\begin{figure}[ht]
		\begin{center}
			\includegraphics[width=0.45\textwidth]{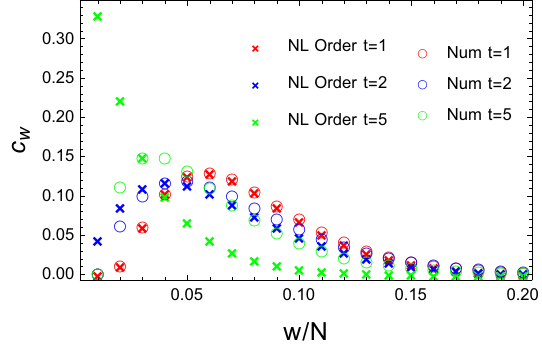}
			\includegraphics[width=0.45\textwidth]{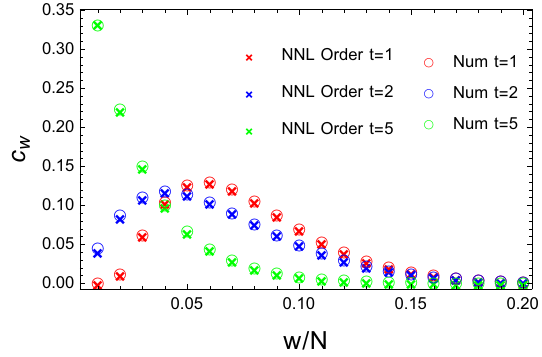}
			\caption{Comparison between perturbation theory and numerical simulation for an initial operator of weight $w_0 = 3$. Parameters: $N=100$, $\kappa=0.5$, $r=1$, $a_2=1$, and $a_{j>2}=0$.  }
			\label{fig:cw}
		\end{center}
	\end{figure}

	Besides $\langle w \rangle_c$, one may compute higher moments of the operator size by taking higher derivatives of the generating function; such moments are relevant for evaluating higher-order corrections to the OTOC. Furthermore, expanding the generating function yields the full distribution $c_w$ of operator weights. As shown in Fig.~\ref{fig:cw}, for an initial weight $w = 3$, at least second-order perturbation theory is required to obtain a distribution that agrees well with numerical simulations.

	\section{Higher-order effects: Three-body interactions}\label{sec:L3}
	In this section, we analyze higher-order corrections for the three-body interaction case. While the perturbative structure is formally similar to the two-body case, three-body interactions introduce a qualitatively new feature: operator growth is constrained by a parity selection rule, which fundamentally alters the late-time dynamics.
	
	As stated in Section~\ref{sec:eq}, three-body interactions change the operator weight by $\Delta w = 0\,,\,\pm 2\,$. As a result, even and odd weight sectors are dynamically decoupled. This constraint leads to multiple long-lived metastable states and makes the role of higher-order corrections even more pronounced than in the two-body case.
	
	Starting from the exact evolution equation for $b_w$
	\begin{align}
		\frac{db_w}{dt}=4\mu_3\left[ \sum_{p\  \text{odd},p+m\le3}rC_{3,p,m}^{w+2m+p-3}b_{w+2m+p-3}(t)-C_{3}^{w}b_w(t)\right]-2\kappa wb_w \;,
	\end{align}
	where we choose $\mu_3^2=\frac{a_3}{18N^2}$. Without lose of generality, we can set $a_3=1$, finally we obtain
	\begin{equation}\label{eq:L3-bw}
		\begin{aligned}
			\partial_t b_w&=2r(w-2)b_{w-2}-2(\kappa+1)wb_{w}\\
			&\newline+\frac{1}{N}\left[\frac{2}{3}wb_{w}(2r(w-1)+4w+5)-2r\left(2w^{2}-7w+6\right)b_{w-2}\right]\\
			&\newline+\frac{1}{N^{2}}\bigg[-\frac{4}{27}wb_{w}\left(r\left(7w^{2}-3w-4\right)+8w^{2}+12w+7\right)\\
			&\newline\newline+\frac{2}{9}rw\left(w^{2}+3w+2\right)b_{w+2}+2r(w-1)(w-2)^{2}b_{w-2}\bigg] \;.
		\end{aligned}
	\end{equation}
	It then follows straightforwardly that the evolution of $\mathsf{G}(x,t)$ is governed by
	\begin{equation}
		\partial_t \mathsf{G}(x, t) = \left(\widehat{A}_0+\frac{\widehat{A}_1}{N}+\frac{\widehat{A}_2}{N^2}\right)\mathsf{G}(x,t) \;,
	\end{equation}
	where
	\begin{equation}
		\begin{aligned}
			\widehat{A}_0&=2rx^{3}\partial_{x}-2(1+\kappa)x\partial_{x} \;,\\
			\widehat{A}_1&= -2rx^{2}(3x\partial_{x}+2x^{2}\partial_{x}^{2})+6x\partial_{x}+\frac{4}{3}(r+2)x^{2}\partial_{x}^{2} \;,\\
			\widehat{A}_2&=-4x\partial_{x}-\frac{4}{27}(18(r+2)x^{2}\partial_{x}^{2}+(7r+8)x^{3}\partial_{x}^{3})\\
			&\newline+2rx^{2}(2x\partial_{x}+4x^{2}\partial_{x}^{2}+x^{3}\partial_{x}^{3})+\frac{1}{x^{2}}\frac{2r}{9}x^{3}\partial_{x}^{3} \;.
		\end{aligned}
	\end{equation}
	In Section~\ref{sec:23-body}, the eigenfunctions of the zeroth-order operator $\widehat{A}_0$ have been derived as $G_k^{(0)}(x) = \frac{x^k}{(1-\frac{\kappa r}{1+\kappa}x^2)^{k/2}}$. We perform the first-order perturbation analysis as
	\begin{align}
		G_k^{(1)}(x)&=\left(-\frac{ \left(3 \left(\kappa ^2-1\right)+2 r^2 x^2-3 (\kappa +1) r+4 r x^2\right)k^2}{6 \left(\kappa -r x^2+1\right)^2}\right. \nonumber \\
		&\quad \left.-\frac{ \left(-3 \left(\kappa ^2+5 \kappa +4\right)-2 r^2 x^2+(9 \kappa +14) r x^2\right)k}{6 \left(\kappa -r x^2+1\right)^2}+C_k^{(1)}\right)G^{(0)}_k(x) \;,\\
		\lambda_k^{(1)}&=\frac{2}{3}  (2 r+4)k^2+\frac{2}{3} (5-2 r)k,~C_{k}^{(1)}=-\frac{k(\kappa+k(-\kappa+r+1)+4)}{2(\kappa+1)}\;.
	\end{align}
	Consistent with the procedure in the two-body case, we substitute $k^n$ with the operator $(\widehat{A}_0 / \lambda_1^{(0)})^n$. This allows us to define the operators $\widehat{O}^{(1)}_x$ and $\widehat{\Lambda}^{(1)}_x$ such that $G^{(1)}_k(x) = \widehat{O}^{(1)}_x G^{(0)}_k(x)$ and $\lambda_k^{(1)} G^{(0)}_k(x) = \widehat{\Lambda}^{(1)}_x G^{(0)}_k(x)$. Using these operators, together with the characteristic flow $x_t$ derived in Section~\ref{sec:23-body}, the first-order corrections to the generating function can be computed following exactly the same procedure as in the two-body case. Subsequently, we present the results for the second-order perturbation as
	\begin{equation}
		\begin{aligned}
			G^{(2)}_k(x)&=\left(g_1(x)k+g_2(x)k^2+g_3(x)k^3+g_4(x)k^4+C_k^{(2)}\right)G_k^{(0)}(x) \;,\\
			\lambda_k^{(2)}&=B_{3}^{(2)}k^{3}+B_{2}^{(2)}k^{2}+B_{1}^{(2)}k \;,
		\end{aligned}
	\end{equation}
	where the functions $g_i(x)$ are provided in Appendix~\ref{appdix}. Since the coefficient $C_k^{(2)}$ does not affect the second-order corrections, its explicit form is omitted here.
	
	By replacing $k^n$ with the operator $(\widehat{A}_0 / \lambda_1^{(0)})^n$, we define the operators $\widehat{O}_x^{(2)}$ and $\widehat{\Lambda}_x^{(2)}$ such that $G^{(2)}_k(x) = \widehat{O}_x^{(2)} G_k^{(0)}(x)$ and $\lambda_k^{(2)} = \widehat{\Lambda}_x^{(2)} G_k^{(0)}(x)$. Substituting these into Eq.~\eqref{eq:gen_NNL} derived in the previous section yields the second-order perturbation.
	\begin{figure}[ht]
		\begin{center}
			\includegraphics[width=0.4\textwidth]{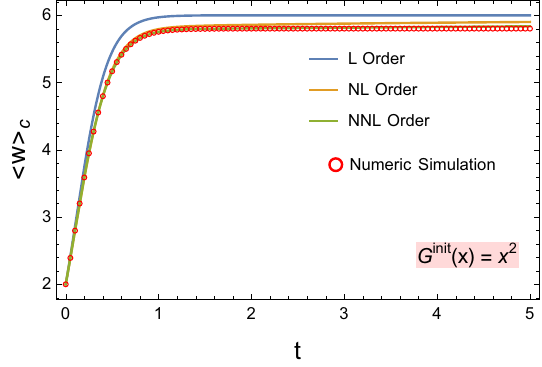}
			\includegraphics[width=0.4\textwidth]{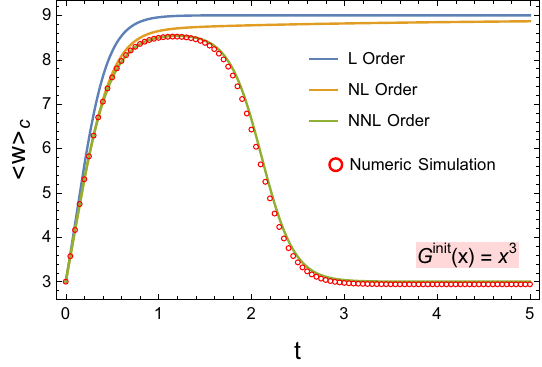}
			\caption{Comparison of perturbative results with numerical simulations of Eq.~\eqref{eq:L3-bw} for three-body interactions. Parameters are set to $N=100$, $\kappa=0.5$, $r=1$, $a_3=1$, and $a_{j \neq 3}=0$. }
			\label{fig:L3-wc}
		\end{center}
	\end{figure} 
	To validate our approach, we compare in Fig.~\ref{fig:L3-wc} the numerical results with analytical predictions obtained from perturbation theory up to second order. We find that, depending on the parity of the initial operator weight $w_0$, the late-time plateau of the average size $\langle w \rangle_c$ takes two distinct values: the plateau for even $w_0$ is approximately twice as high as that for odd $w_0$.
	
	This phenomenon can also be understood within the interaction‑picture perturbative framework. Unlike in two‑body interactions, the perturbative term $M_\infty^I$ in the three‑body case shifts the operator‑weight distribution leftward by \emph{two} steps at a time. If the initial distribution contains only even weights, $M_\infty^I$ can at most shift weight to $w=2$. Because a purely even distribution has zero overlap with the ``ground state" (whose eigenvalue has the largest real part) of the unperturbed ``Hamiltonian" $M_\infty^{(0)}$, the late-time operator distribution is given by $G_2^{(0)}(x)$, and the $\langle w\rangle_c$ approaches the plateau $2/(1-r_{\text{eff}})+\mathcal{O}(1/N)$.
	
	Conversely, if the initial distribution contains an odd component, $M_\infty^{I}$ can eventually shift weight to $w=1$. The state with $b_1\not=0$ has non‑zero overlap with the ground state of $M_\infty^{(0)}$; therefore, in the long‑time limit, the ground‑state contribution $G_1^{(0)}(x)$ dominates and the system relaxes to the lower plateau $1/(1-r_{\text{eff}})+\mathcal{O}(1/N)$. This mechanism implies that for a mixed‑parity initial distribution, the odd component ultimately determines the steady‑state behavior, yielding the same plateau as that of a purely odd initial condition.

    \section{Discussion}
    \label{sec:discussion}
    
    In this work, we developed a systematic framework for analyzing operator growth beyond leading order in Brownian spin models, with particular emphasis on the dilute regime where operator weights remain parametrically smaller than the system size. By extending the standard Brownian circuit approach to incorporate higher-order corrections and decoherence effects, we obtained a closed and analytically tractable description of the time evolution of the full operator-size distribution $b_w(t)$. The generating-function formulation provides a unifying language for these dynamics: it allows arbitrary initial operator-size distributions to be evolved analytically and separates model-dependent ingredients, such as the interaction order and decoherence rate, from the general structural properties of the evolution.
    
    A central conceptual outcome of our analysis is that higher-order corrections in $1/N$ are not merely quantitative refinements. At leading order in the dilute limit, the evolution matrix governing $b_w$ becomes lower triangular, leading to a hierarchy of decay rates that can be diagonalized explicitly. However, this triangular structure is a consequence of the leading-order truncation. Once subleading corrections are included, different operator-weight sectors become weakly coupled, and this coupling is essential for the correct late-time behavior. In particular, higher-order terms generate the mode mixing needed for the operator-size distribution to relax toward its universal attractor. This explains why, for an initial operator of weight $m$, corrections up to order $m-1$ are generally required to capture the asymptotic relaxation.
    
    The generating-function method used in this work relies on extending the finite transfer matrix to an infinite-dimensional one while retaining its explicit $N$-dependence. This step is natural within the dilute regime, where the average operator size remains much smaller than $N$, but it also introduces an approximation whose rigorous error estimate is not provided here. Numerically, we expect that in the dilute regime the error associated with this infinite-dimensional extension is strongly suppressed at large $N$, while the perturbative error from the $1/N$ expansion is only algebraically suppressed. A more systematic mathematical justification of this point would be desirable. The validity of the method beyond the dilute regime, where saturation and finite-$N$ effects become important, remains an open question.
    
    It is useful to contrast the present generating-function approach with a direct interaction-picture treatment of the finite-dimensional transfer matrix. In principle, the $1/N$ perturbative expansion can be formulated without extending the matrix dimension, by inserting subleading interaction terms into the leading-order evolution operator. Such a formulation would avoid the infinite-dimensional approximation, but it typically involves repeated finite-matrix multiplications and becomes technically cumbersome. The generating-function method effectively converts these matrix operations into differential-operator manipulations, which is the main source of its calculational simplicity. Away from the dilute regime, however, a direct interaction-picture approach may provide a more robust complementary method.
    
    The present formalism should also be distinguished from the continuum approximation often used in operator-growth problems. In the continuum approximation, the discrete operator weight $w$ is treated as a continuous coordinate, and weight-shift terms in the evolution equation are expanded in derivatives with respect to this coordinate. This approximation is useful when the operator-size distribution is broad and sufficiently smooth. By contrast, in our generating-function formalism, the variable $x$ is only an auxiliary bookkeeping parameter and has no direct interpretation as a physical spatial coordinate. The discreteness of the operator weight is retained, while the dynamics is encoded in a differential equation for the generating function.
    
    There are several natural directions for future work. One possibility is to apply related perturbative ideas to other diagnostics of operator growth, such as Krylov complexity. The evolution equation for the Krylov amplitudes has a structure similar to the master equation studied here, but the corresponding transfer matrix is generally not lower triangular and is usually analyzed using orthogonal-polynomial methods. It would be interesting to understand whether a suitably modified generating-function perturbation theory can provide useful analytic control over Krylov-complexity growth beyond known solvable cases.
    
    Another important question is how far the present approach can be generalized beyond all-to-all Brownian models. In the present work, the all-to-all structure plays an important technical role: because of permutation symmetry, the ensemble-averaged dynamics closes on the total operator weight $w$, leading to a one-dimensional master equation and a single-variable generating function. For spatially local systems, such as local random circuits or deterministic spin chains, the total operator weight alone is generally insufficient, since the spatial profile of the operator and the structure of the operator front also become important. A direct extension would therefore require spatially resolved operator-size distributions, or equivalently multi-variable generating functions. Whether the parity-protected or sectorized metastable dynamics found here survives in such spatially structured settings is an interesting problem for future investigation.
    
    A related extension concerns operator-space entanglement entropy. The present work focuses on the operator-size distribution, which keeps track of the total Pauli weight but discards the spatial arrangement of the Pauli strings. Therefore, the single-variable distribution $b_w(t)$ is not sufficient to determine operator entanglement across a bipartition. A possible generalization is to introduce a joint size distribution $b_{w_A,w_B}(t)$, where $w_A$ and $w_B$ denote the Pauli weights in the two subsystems. For Brownian all-to-all models, the remaining permutation symmetry within each subsystem may still allow a closed evolution equation for this joint distribution, together with a corresponding two-variable generating function. Such a framework could provide access to operator-space R\'enyi entropies and clarify the relation between operator-size growth and operator-space entanglement \cite{Prosen2007,Prosen2007-2,Pizorn:2009gup}. We leave a detailed analysis of this direction for future work.
    
    Finally, it is tempting to connect the suppression of operator growth observed in our model to the operator-size/momentum correspondence \cite{susskind2018thingsfall}. In this dual picture, the saturation of operator size is related to the slowing down of the radial momentum of an infalling particle near a black-hole horizon \cite{Brown_2018,Ageev_2019}. From this perspective, the non-ideal parameters introduced in our setup may be viewed heuristically as producing an effective damping of operator growth. We emphasize, however, that this analogy remains speculative, and a precise holographic interpretation of Brownian operator growth with imperfections and decoherence remains an open problem.
    
	\section*{Acknowledgments}
	We thank  Cheng Peng for useful discussions in the project. 
	This work is supported by
	NSFC NO. 12175237, and NSFC NO. 12447108, the Fundamental Research Funds for the
	Central Universities, and funds from the Chinese Academy of Sciences.
	\appendix
	\section{Solution of NNL-order eigenfunctions for two-body and three-body interactions}
	\label{appdix}
	\allowdisplaybreaks
	\subsection{Two-body interactions}
	\begin{align}
		\lambda_{k}^{(2)}&=\frac{2kr^{2}\left(-3\kappa^{2}-5\kappa+4k^{2}\left(-3\kappa^{2}-2\kappa+3r^{2}+1\right)+k\left(9\kappa^{2}+15\kappa-6r^{2}+6\right)+2r^{2}-2\right)}{9(\kappa+1)^{3}}\ed
	\end{align}
	\begin{align}
		G^{(2)}_k(x)&=-\frac{x^k (\kappa -r x+1)^{-k}}{18 (\kappa +1)^2} \left(-\frac{k (k+1) \left(k^2 \left(4 (\kappa +1)^2+3 r^4+3 (\kappa +1) (\kappa +3) r^2\right)\right)}{(\kappa -r x+1)^2}\right. \nonumber \\
		&\quad \left.-\frac{k(k+1)\left(3 k \left(8 (\kappa +1)^3+r^4+(\kappa +1) (5 \kappa +7) r^2\right)+(\kappa +1)^2 \left(45 \kappa ^2+84 \kappa -6 r^2+38\right)\right)}{(\kappa -r x+1)^2}\right. \nonumber \\
		&\quad \left.+\frac{k \left(-6 (\kappa +1)^3 (3 \kappa +4)+k^3 r^2 \left(3 \kappa ^2-5 r^2-3\right)+2 k^2 r^2 \left(10 r^2-(\kappa +1) (15 \kappa +2)\right)\right)}{(\kappa +1) (-\kappa +r x-1)}\right. \nonumber \\
		&\quad \left.+\frac{k^2 \left(-12 (\kappa +1)^3+r^4+(\kappa +1) (3 \kappa +11) r^2\right)}{(\kappa +1) (-\kappa +r x-1)}-\frac{(k-2) (k-1)^2 k r^2}{x^2}\right. \nonumber \\
		&\quad \left.+\frac{(k-1) k r \left(k \left(3 (\kappa +1) (\kappa +5)+k \left(-3 \kappa ^2+5 r^2+3\right)+9 r^2\right)-2 \left(\kappa +2 r^2+1\right)\right)}{(\kappa +1) x}\right. \nonumber \\
		&\quad \left.+\frac{(\kappa +1) k (k+1) (k+2) \left(r^2-(\kappa +1) (3 \kappa +1)\right) \left((\kappa +1) (21 \kappa +20)+6 k \left(\kappa +r^2+1\right)\right)}{3 (-\kappa +r x-1)^3}\right. \nonumber \\
		&\quad \left.-\frac{(\kappa +1)^2 k (k+1) (k+2) (k+3) \left(r^2-(\kappa +1) (3 \kappa +1)\right)^2}{4 (\kappa -r x+1)^4}\right)+C_k^{(2)}G_k^{(0)}(x) \;.
	\end{align}
	\subsection{Three-body interactions}
	\begin{align}
		\lambda_k^{(2)}&=\frac{2 \left(-16 \kappa +12 r^2-14 \kappa  r-14 r-16\right)k^3 }{27 (\kappa +1)}+\frac{2 \left(-24 \kappa -9 r^2+6 \kappa  r+6 r-24\right) k^2}{27 (\kappa +1)}\nonumber\\ &\quad+\frac{2  \left(-14 \kappa +6 r^2+8 \kappa  r+8 r-14\right) k}{27 (\kappa +1)}\ed
	\end{align}
	\begin{align}
		g_1(x) &=\frac{1}{648}  \left(\frac{180 r^2}{\left(\kappa -r x^2+1\right)^2}+\frac{9936 \kappa ^2 (\kappa +1)}{\left(-\kappa +r x^2-1\right)^3}+\frac{64 (\kappa +1) (9 r-59) r}{\left(-\kappa +r x^2-1\right)^3}-\frac{5168 \kappa  (\kappa +1) r}{\left(-\kappa +r x^2-1\right)^3}\right. \nonumber \\
		&\quad \left.-\frac{72 r}{(\kappa +1) x^2}-\frac{48 (68 \kappa +77) r}{\left(\kappa -r x^2+1\right)^2}+\frac{432 (\kappa +1)^2 (-3 \kappa +r-1)^2}{\left(\kappa -r x^2+1\right)^4}+\frac{13040 \kappa  (\kappa +1)}{\left(-\kappa +r x^2-1\right)^3}\right. \nonumber \\
		&\quad \left.+\frac{3200 (\kappa +1)}{\left(-\kappa +r x^2-1\right)^3}+\frac{6 \kappa  (1377 \kappa +2680)}{\left(\kappa -r x^2+1\right)^2}-\frac{12 (-189 \kappa +62 r-338)}{-\kappa +r x^2-1}+\frac{7548}{\left(\kappa -r x^2+1\right)^2}\right) \;,
	\end{align}
	\begin{align}
		g_2(x) &= \frac{1}{648} \left(\frac{90 r^2}{\left(\kappa -r x^2+1\right)^2}+\frac{7452 \kappa ^2 (\kappa +1)}{\left(-\kappa +r x^2-1\right)^3}+\frac{48 (\kappa +1) (9 r-59) r}{\left(-\kappa +r x^2-1\right)^3}+\frac{24 r (56 \kappa -9 r+35)}{\left(\kappa -r x^2+1\right)^2}\right. \nonumber \\
		&\quad \left.-\frac{3876 \kappa  (\kappa +1) r}{\left(-\kappa +r x^2-1\right)^3}+\frac{108 r}{(\kappa +1) x^2}-\frac{24 (68 \kappa +77) r}{\left(\kappa -r x^2+1\right)^2}+\frac{396 (\kappa +1)^2 (-3 \kappa +r-1)^2}{\left(\kappa -r x^2+1\right)^4}\right. \nonumber
		\\
		&\quad \left.+\frac{9780 \kappa  (\kappa +1)}{\left(-\kappa +r x^2-1\right)^3}+\frac{2400 (\kappa +1)}{\left(-\kappa +r x^2-1\right)^3}+\frac{48 (59 \kappa +62)}{\left(\kappa -r x^2+1\right)^2}+\right. \nonumber \\
		&\quad \left.\frac{3 \kappa  (1377 \kappa +2680)}{\left(\kappa -r x^2+1\right)^2}+\frac{144 (4 r+11)}{-\kappa +r x^2-1}-\frac{288 (\kappa +1) (r+2) (-3 \kappa +r-1)}{\left(-\kappa +r x^2-1\right)^3}+\frac{3774}{\left(\kappa -r x^2+1\right)^2}\right) \;,
	\end{align}
	\begin{align}
		g_3(x) &= \frac{1}{648} \left(\frac{12 \left(16 (\kappa +1)-9 r^2+14 (\kappa +1) r\right)}{(\kappa +1) \left(-\kappa +r x^2-1\right)}+\frac{1242 \kappa ^2 (\kappa +1)}{\left(-\kappa +r x^2-1\right)^3}+\frac{72 (r+2)^2}{\left(\kappa -r x^2+1\right)^2}\right. \nonumber \\
		&\quad \left.-\frac{216 (\kappa +1) (r+2) (-3 \kappa +r-1)}{\left(-\kappa +r x^2-1\right)^3}+\frac{108 (\kappa +1)^2 (-3 \kappa +r-1)^2}{\left(\kappa -r x^2+1\right)^4}+\frac{1630 \kappa  (\kappa +1)}{\left(-\kappa +r x^2-1\right)^3}\right. \nonumber \\
		&\quad \left.+\frac{8 (\kappa +1) r (9 r-59)}{\left(-\kappa +r x^2-1\right)^3}+\frac{400 (\kappa +1)}{\left(-\kappa +r x^2-1\right)^3}+\frac{12 r (56 \kappa -9 r+35)}{\left(\kappa -r x^2+1\right)^2}+\frac{24 (59 \kappa +62)}{\left(\kappa -r x^2+1\right)^2}\right. \nonumber \\
		&\quad \left.-\frac{646 \kappa  (\kappa +1) r}{\left(-\kappa +r x^2-1\right)^3}-\frac{36 r}{(\kappa +1) x^2}\right) \;,
	\end{align}
	\begin{align}
		g_4(x)&=\frac{1}{648}\left(\frac{36 (r+2)^2}{\left(\kappa -r x^2+1\right)^2}-\frac{36 (\kappa +1) (r+2) (-3 \kappa +r-1)}{\left(-\kappa +r x^2-1\right)^3}+\frac{9 (\kappa +1)^2 (-3 \kappa +r-1)^2}{\left(\kappa -r x^2+1\right)^4}\right)\;.
	\end{align}
	
	\bibliographystyle{JHEP}
	\bibliography{ref2}
\end{document}